\documentclass[a4paper,fleqn,usenatbib]{mnras}
\usepackage[T1]{fontenc}
\usepackage{ae,aecompl}
\usepackage{amsmath}
\usepackage{amsfonts}
\usepackage{amssymb}
\usepackage{graphicx}
\usepackage{color,xcolor}
\def\mnras{MNRAS}
\def\aap{A$\&$A}
\def\nat{Nature}
\def\apj{ApJ}
\def\apjl{ApJ Letters}
\def\apjs{ApJS}

\def\aj{AJ}
\def\icarus{Icarus}

\newcommand{\me}{\, {\rm M}_{\oplus}}
\newcommand{\msun}{\, {\rm M}_{\odot}}

\newcommand{\rsun}{\, {\rm R}_{\odot}}
\newcommand{\au}{\, {\rm au}}
\newcommand{\ab}{\, a_{\rm bin}}

\newcommand{\acav}{\, {r_{\rm cav}}}
\newcommand{\ecav}{\, {e_{\rm cav}}}
\newcommand{\rcavapo}{\, {r_{\rm cav,a}}}

\title[Properties of Free Floaters]{On the properties of free floating planets originating in circumbinary planetary systems}
\author[G. A. L. Coleman]{Gavin A. L. Coleman\thanks{Email: gavin.coleman@qmul.ac.uk}\\
Astronomy Unit, Department of Physics and Astronomy, Queen Mary University of London, Mile End Road, London, E1 4NS, UK}
\date{Accepted 2024 March 26; Received 2024 March 25; in original form 2024 February 8}
\pubyear{2024}
\begin{document}
\label{firstpage}
\pagerange{\pageref{firstpage}--\pageref{lastpage}}
\maketitle
\begin{abstract}
Free-floating planets are a new class of planets recently discovered. These planets don't orbit within stellar systems, instead living a nomadic life within the galaxy. How such objects formed remains elusive. Numerous works have explored mechanisms to form such objects, but have not yet provided predictions on their distributions that could differentiate between formation mechanisms. In this work we form these objects within circumbinary systems, where these planets are readily formed and ejected through interactions with the central binary stars. We find significant differences between planets ejected through planet-planet interactions and those by the binary stars. The main differences that arise are in the distributions of excess velocity, where binary stars eject planets with faster velocities. These differences should be observable amongst known free-floating planets in nearby star-forming regions. We predict that targeted observations of directly imaged free-floating planets in these regions should be able to determine their preferred formation pathway, either by planet formation in single or multiple stellar systems, or through processes akin to star formation. Additionally the mass distributions of free-floating planets can yield important insights into the underlying planet populations. We find that for planets more massive than 20$\me$, their frequencies are similar to those planets remaining bound and orbiting near the central binaries. This similarity allows for effective and informative comparisons between mass distributions from microlensing surveys, to those of transit and radial velocities. Ultimately, by observing the velocity dispersion and mass distribution of free-floating planets, it will be possible to effectively compare with predictions from planet formation models, and to further understand the formation and evolution of these exotic worlds.
\end{abstract}

\section{Introduction}
\label{sec:intro}

Until recently planets have only been observed orbiting either a single star \citep[see][for a review]{Winn15} or pairs of stars (e.g. Kepler-16b \citep{Doyle11} and recently BEBOP-1c \citep{Standing23}).
In the last decade however, microlensing surveys have discovered a handful of planets that are not bound to any known stellar system \citep{Mroz19,Mroz20,Koshimoto23,Sumi23}.
Such ``free-floating planets'' (FFPs) have now been found to exhibit a variety of planet masses, ranging from terrestrial \citep{Mroz20}, to Neptunian \citep{Mroz19,Koshimoto23} to Jovian \citep{Sumi11}.
Additionally, direct imaging of nearby clusters have also discovered numerous planetary mass objects that are not attached to close-by systems, whilst also showing evidence of planetary multiplicity \citep{Pearson23}.
With more of these planets now being discovered, their formation remains an interesting and unsolved question.

There are several mechanisms that can explain the formation of FFPs, ranging from those derived from star formation processes, and those from planet formation.
From star formation routes, FFPs can arise through: core collapse models \citep{Padoan02,MiretRoig22,PortegiesZwart24}, aborted stellar embryo ejection from stellar nurseries \citep{Reipurth01} and through photo-erosion of a pre-stellar core by energetic stellar winds \citep{Whitworth04}. From planet formation scenarios, FFPs are formed similarly to other planets, but may undergo planet--planet scattering \citep{RasioFord1996,Weidenschilling96,Veras12}, can be ejected through stellar flybys \citep{Wang24}, or they may be ejected through interactions with binary stars in circumbinary systems \citep{Nelson03,Sutherland16,Smullen16,Coleman23,Coleman24,Standing23,Cheng24}. It is worth noting that planet--planet scattering around single stars cannot explain the large number of FFPs seen in observations \citep{Veras12,Ma16}.

Whilst the mechanisms for the formation of FFPs are numerous, those involving planet formation have not explored the formation of FFPs from full population models, including the growth and evolution of the planetary systems. Additionally they generally focus on specific types of planet, e.g. giant planets, or by demonstrating efficient mechanisms for FFP formation. Previous global models of circumbinary planet formation showed that planets were frequently ejected from the systems, as they formed and grew in the circumbinary disc \citep{Standing23,Coleman23,Coleman24}. These outcomes were all natural byproducts of the planet formation process where the models were able to form planetary systems similar to Kepler-16, -34, and BEBOP-1.

In this paper, we utilise these models and run numerous simulations exploring the effects of different initial conditions on the planet formation process, and the final planetary systems. We focus on the formation and ejection of FFPs that arise naturally in the simulations, and find numerous trends within the distributions of planet properties. These include being able to differentiate between whether planets are ejected by the binary stars or through planet--planet interactions, and also the effects of different parameters on the mass, number and velocity distributions of FFPs. The distributions and predictions derived from our results will provide useful comparisons with present and future observations of FFPs, both those found through microlensing surveys finding old and evolved planets \citep{Sumi23}, and additionally those found through direct imaging within nearby clusters \citep{Pearson23}.

This paper is laid out as follows.
We outline our physical model in Sect. \ref{sec:base_model}, whilst we describe our population parameters in Sect. \ref{sec:pop_parameters}.
In Sect. \ref{sec:results}, we outline the results of our population models, whilst in Sect. \ref{sec:parameter_effects} we explore the effects that changing initial conditions have on the distributions of FFPs. 
Finally, we discuss our results and draw conclusions in Sect. \ref{sec:conc}.

\section{Physical Model}
\label{sec:base_model}

In the following section, we provide a basic overview of the physical model we adopt and the numerical scheme used to undertake the simulations.
The N-body simulations presented here were performed using the {\sc{mercury6}} symplectic N-body integrator \citep{Chambers}, updated to accurately model planetary orbits around a pair of binary stars \citep{ChambersBinary}.
We utilise the `close-binary' algorithm described in \citet{ChambersBinary} that calculates the temporal evolution of the positions and velocities of each body in the simulations with respect to the centre of mass of the binary stars, subject to gravitational perturbations from both stars and other large bodies.
We also include prescriptions for the evolution of 1D protoplanetary discs as well as disc-planet interactions.
With the disc models being 1D in nature, we also include prescriptions that take into account non-axisymmetric effects (i.e. a precessing eccentric inner disc cavity) due to the binary stars.
Details on the full model and the additional prescriptions due to the binary can be found in \citet{Coleman23,Coleman24}. However we briefly describe the model below.

(i) We solve the standard diffusion equation for a 1D viscous $\alpha$-disc model \citep{Shak,Lynden-BellPringle1974}.
Disc temperatures are calculated by balancing black-body cooling against viscous heating and stellar irradiation from both stars.
The viscous parameter $\alpha$ remains constant throughout most of the disc, but increases close to the central stars to mimic the eccentric cavity that is carved out by the tidal forces of the central stars.
When giant planets are present, tidal torques from the planets are applied to the disc leading to the opening of gaps \citep{LinPapaloizou86}.

(ii) We incorporate models of photoevaporative winds removing material from the disc, both internally driven through radiation emanating from the central stars, and externally driven due to FUV radiation from nearby sources (i.e. O/B--stars).
For internal photoevaporation, mainly due to EUV and X-ray radiation from the central stars \citep{Clarke2001,Owen10}, we include the photoevaporation models of \citep{Ercolano21,Picogna21} where the wind is assumed to be launched thermally from the disc upper and lower surfaces beyond a critical radius. We assume the most massive star dominates the high-energy radiation driving internal photoevaporation, and assume it has an X-ray Luminosity $L_{\rm X}=10^{30.5}$, consistent with the aveage value for Solar mass stars in star forming regions \citep{Flaischlen21}.
With ionising radiation not just impacting the disc from the central stars, but also from nearby stars in the local star-forming region \citep[e.g.][]{Haworth18,Haworth23}, we include external photoevaporation in the models to account for the effects of ionising FUV photons.
We adopt the model found in \citep{Qiao23} which drives a wind from outside the radius where the disc becomes optically thin.

(iii) Eccentric cavities, arising because of the tidal torque from the central binary, have been seen in observations and numerical simulations of circumbinary discs \citep[e.g.][]{Artymowicz94,Dutrey94,Pierens13,Mutter17D,Thun17,Coleman22b}.
The shape and size of these cavities depends on the binary properties, i.e. mass ratio and eccentricity, and local disc properties such as the viscosity parameter $\alpha$ \citep{Kley19}.
The main effect of circumbinary disc cavities on planet formation is the creation of a planet migration trap as the corotation torque is increased due to the positive surface density gradient at the cavity edge.
To simulate the effects of an eccentric cavity in our 1D disc model, we ran 2D hydrodynamical simulations of circumbinary discs using \textsc{fargo3d} \citep{FARGO-3D-2016} to determine the cavity structure. In the 1D models, we simulate the azimuthally averaged surface density profile of the cavity by adjusting the viscosity parameter $\alpha$, whilst maintaining a constant gas flow rate through disc. This forms an inner cavity in the disc and leads to a buildup of material at the outer edge of the cavity, as required.

(iv) Using the above 2D hydrodynamical simulations, we take into account the precessing, eccentric nature of the inner disc cavity in 1D models through construction of 2D maps of the gravitational acceleration experienced by test particles embedded in the disc due to the non-axisymmetric density distribution. We also create maps of the gas surface densities and gas velocities.
These maps are used when integrating the equations of motion of planets and when calculating relative velocities between planets and drifting pebbles. These effects are mainly relevant near the cavity region.

(v) We include the planetesimal and embryo formation models found in \citet{Coleman21}. We assume that as the pebble production front moves outward and pebbles drift inward, they can collect in short-lived traps due to non-axisymmetric perturbations in the discs.
As solids collect in the traps, then planetesimals may form through gravitational collapse, when the local particle density exceeds the Roche density, and when the local dust-to-gas ratio exceeds unity \citep{Johansen07,JohansenYoudin2009}.
The size distribution of the planetesimals then follows a power law plus an exponential decay \citep{Johansen15,Schafer17,Abod19}.
We assume that the most massive planetesimal that forms in each collapse event is ultimately a planetary embryo that is able to dominate the accretion and dynamical evolution of other planetesimals that formed in that vicinity.
This planet is then able to accrete pebbles that are drifting past its orbit \citep{Johansen17}, and planetesimals from the newly formed local reservoir \citep{Fortier13}, whilst also undergoing mutual interactions and collisions with other planetary embryos.

(vi) The main source of the accretion of solids in our models is through pebble accretion. We follow the pebble accretion model of \citet{Johansen17}, where a pebble production front moves outwards in the disc over time. This production front arises from dust particles coagulating and settling to the disc midplane forming pebbles. Once these pebbles become large enough, they begin to drift inwards through gas drag forces, thus creating a pebble production front when the drift time-scale is equal to the growth time-scale.
As the pebbles drift inwards they can be accreted by planetary embryos, allowing them to grow on short time-scales \citep{Lambrechts12}.
The accretion of pebbles continues until the planets reach the pebble isolation mass, that being the mass where planets are able to sufficiently perturb the local disc, forming a pressure bump exterior to the planet's orbit, that traps pebbles and halts pebble accretion on to the planet \citep{Lambrechts14,Ataiee18,Bitsch18}.

(vii) The accretion of gaseous envelopes on to solid cores occurs once a planet’s mass exceeds 1$\me$.
We utilise the formulae based in \citet{Poon21} that are based on fits to gas accretion rates obtained using a 1D envelope structure model \citep{Pap-Terquem-envelopes,PapNelson2005,CPN17}.
To calculate these fits \citet{Poon21} performed numerous simulations, embedding planets with initial core masses between 2--15 $\me$ at orbital radii spanning 0.2--50 $\au$, within gas discs of different masses.
This allowed for the effects of varying local disc properties to be taken into account when calculating fits to gas accretion rates, a significant improvement on fits used in previous work \citep[e.g.][]{ColemanNelson14,ColemanNelson16,ColemanNelson16b}.
We use these fits until a planet is massive enough to undergo runaway gas accretion and open a gap in the disc.
The gas accretion rate is then limited to either the maximum value of the fits from \citet{Poon21}, or the viscous supply rate.
All gas that is accreted onto a planet is removed from the surrounding disc, such that the accretion scheme conserves mass.

(viii) We use the torque formulae from \citet{pdk10,pdk11} to simulate type I migration due to Lindblad and corotation torques acting on planetary embryos.
Corotation torques arise from both entropy and vortensity gradients in the disc, and the possible saturation of these torques is included in the simulations.
The influence of eccentricity and inclination on the migration torques, and of eccentricity and inclination damping are included \citep{Fendyke,cressnels}.

(ix) Type II migration of gap forming planets is simulated using the impulse approximation of \citet{LinPapaloizou86}, where we use the gap opening criterion of \citet{Crida} to determine when to switch between type I and II migration.
Thus, when a planet is in the gap opening regime, the planet exerts tidal torques on the disc to open a gap, and the disc back-reacts onto the planet to drive type II migration in a self-consistent manner.

\begin{table}
\centering
\begin{tabular}{l|lc}
Parameter & Description & Value\\
\hline
$M_{\rm A}\ (\msun)$ & Primary Mass & $1.0378$\\
$M_{\rm B}\ (\msun)$ & Secondary Mass & $0.2974$\\
$T_{\rm A}\ ({\rm K})$ & Primary Temperature & 4300\\
$T_{\rm B}\ ({\rm K})$ & Secondary Temperature & 3300\\
$R_{\rm A}\ (\rsun)$ & Primary Radius & 2\\
$R_{\rm B}\ (\rsun)$ & Secondary Radius & 1.5\\
$e_{\rm bin}$ & Binary Eccentricity & $0.156$\\
Metallicity (dex) & Stellar Metallicity & 0\\
& & \\
$\acav\ (\ab)$ & Cavity Radius & 3.7377\\
$\ecav$ & Cavity Eccentricity & 0.4162\\
$\rcavapo\ (\ab)$ & Cavity Apocentre & 5.2933\\
$C_1$ & Cavity Parameter 1 & 1.1\\
$C_2$ & Cavity Parameter 2 & 0.32\\
$C_3$ & Cavity Parameter 3 & 4.5\\
\end{tabular}
\caption{Simulation common parameters.}
\label{tab:sim_param}
\end{table}

\section{Population Parameters}
\label{sec:pop_parameters}

Previous work showed that ejected planets are a natural outcome of circumbinary planet formation simulations \citep{Standing23,Coleman23,Coleman24}.
Those studies varied the initial disc mass, metallicity, strength of the external environment, and the level of turbulence in the disc.
However their main focus was on the types of circumbinary systems that form, and not on the properties of planets ejected from those systems.
For the population of circumbinary systems in this work, we vary the initial disc mass, the binary separation, the strength of the external environment, and the level of turbulence in the disc.
We only explore a single combined stellar mass, mass ratio and binary eccentricity, since for each combined set of those parameters, the properties of the central cavities are different, and so varying them would require large numbers of hydrodynamic simulations to be run to calculate the properties of the cavities.
We base the central stars on that found in the BEBOP-1 circumbinary system \citep{Kostov20,Standing23}.
The stellar and model parameters used to simulate the central cavities in the circumbinary discs as described above can be found in Table. \ref{tab:sim_param}.
The reason we choose the BEBOP-1 parameters for our simulations, is that the 1D circumbinary disc models require hydrodynamical simulations to be performed in order to provide prescriptions for the circumbinary cavity, disc eccentricity and the multi-dimensional maps for the resulting torques and gas velocities (see points iii and iv in Sect. \ref{sec:base_model}). These simulations were already performed in recent work \citep{Coleman24}. Additionally, choosing the BEBOP-1 parameters allowed preliminary comparisons with previous work that contained some overlapping simulations, but concentrated on different objectives, i.e. forming systems similar to BEBOP-1 \citep{Coleman24}.
The choice of a total combined stellar mass of 1.3$\msun$ is also similar to the average combined stellar mass for binary stars \citep[][$\sim 1.5\msun$]{Raghavan10}.
Table \ref{tab:pop_param} shows the parameters that we do vary, and we take random values between the limits shown in table \ref{tab:pop_param}, with the last column denoting whether we randomise in log or linear space.

We choose initial disc masses between 5 and 15\% of the combined binary mass, with the maximum disc mass being equal to the most massive disc a star can host before it becomes gravitationally unstable \citep{Haworth20}.
Numerous works have provided observational estimates for the viscosity parameter $\alpha$ \citep{Isella09,Andrews10,Pinte16,Flaherty17,Trapman20,Villenave20,Villenave22}, whilst theoretical work has contributed additional constraints \citep{Standing23,Coleman24}.
We adopt values of $\alpha$ that are consistent with such estimates \citep[see][for a recent review]{Rosotti23}.
For the final parameter that affects the disc lifetime, the rate of external photoevaporation, we vary the strength of the local environment UV field, ranging from 1 $\rm G_0$ to $10^{5} \rm G_0$\footnote{$G_0$ is taken as the flux integral over 912--2400\AA, normalised to the value in the solar neighbourhood \citep{Habing68}.}.
These values are consistent with what is expected across star forming regions such as Orion, with other low mass regions such as Taurus and Lupus occupying the lower region of our parameter space \citep{Winter18}.
The final parameter we vary is the binary separation, which we explore separations between 0.05 and 0.5 $\au$. 

\begin{table}
\centering
\begin{tabular}{l|ccc}
Parameter & Lower Value & Upper Value & Dimension\\
\hline
$a_{\rm bin}\ (\au)$ & 0.05 & 0.5 & linear\\
$M_{\rm disc}\ (M_{\rm bin})$ & 0.05 & 0.15 & linear\\
UV Field $(G_0)$ & $1$ & $10^{5}$ & log\\
$\alpha$ & $10^{-4}$ & $10^{-2.5}$ & log\\
\end{tabular}
\caption{Values for the parameters varied amongst the populations.}
\label{tab:pop_param}
\end{table}

We initialise the surface density in the circumbinary disc following \citet{Lynden-BellPringle1974}
\begin{equation}
    \Sigma = \Sigma_0\left(\frac{r}{1\au}\right)^{-1}\exp{\left(-\frac{R}{R_{\rm C}}\right)}
\end{equation}
where $\Sigma_0$ is the normalisation constant set by the total disc mass, (for a given $R_{\rm C}$), and $R_{\rm C}$ is the scale radius, which sets the initial disc size, taken here to be equal to 50 $\au$.
The domain of our circumbinary discs extend from the binary separation as the inner edge to an outer edge of 500 $\au$ where for all discs the surface density is nearly always at our floor value of $10^{-5}\rm g cm^{-2}$.
We run each simulation for 10 Myr to account for the entire circumbinary disc phase, and additionally allowing for dynamical evolution of the systems after the dispersal of the circumbinary discs.
Planets are removed from the simulations once they either collide with the central stars or other planets or when they are ejected from the system.
We define a planet as having been ejected once it enters into a hyperbolic orbit, i.e. having an eccentricity greater than 1, and when its distance from the barycentric centre of the system has exceeded 1000 $\au$.

\section{Results}
\label{sec:results}

The main objectives of this work are to explore the properties and distribution of free-floating planets that originate in circumbinary systems before being ejected.
Whilst we do not discuss the properties of the remaining circumbinary systems themselves, or their formation as a whole, their formation pathways and outcomes are broadly similar to those found in \citet{Coleman24}. We will explore the effects that the varying of the initial parameters in this work have on those outcomes in future work.
Before we explore the distributions of ejected planets, and how the initial conditions affects their properties, we will describe the evolution of a typical simulation that led to multiple ejections from a single circumbinary system.

\begin{figure}
\centering
\includegraphics[scale=0.55]{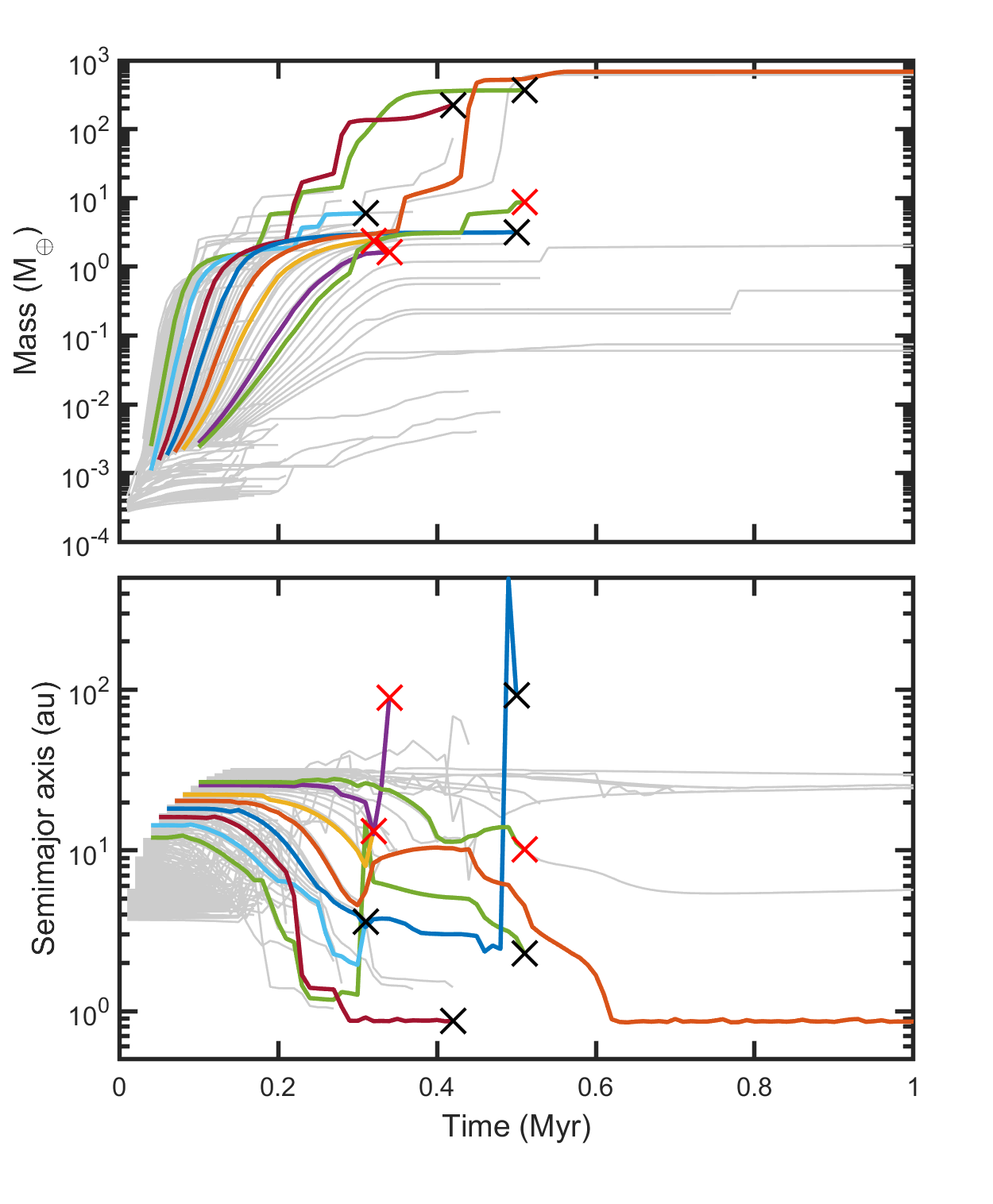}
\caption{Temporal evolution of planet semimajor axes (bottom panel), and masses (top panel) for an example system that ejected multiple planets as described in sect. \ref{sec:example_system}. The coloured lines show those planets that are ejected from the system and with final masses $m_{\rm p}>1\me$. The grey lines show the tracks for all other planets. The black and red crosses show the final semimajor axes and masses of planets ejected through interactions with the central binary, and through planet--planet interactions respectively.}
\label{fig:sim_time}
\end{figure}

\subsection{Example system with multiple ejection sites}
\label{sec:example_system}

The formation pathways of the planets in the circumbinary systems loosely follows that described in previous works \citep[e.g.][]{Coleman23,Coleman24}.
The main difference to those works here, is that by including the planetesimal and embryo formation models of \citet{Coleman21}, the models form the initial planetary embryos, instead of starting with a collection of embryos spread throughout the disc. These embryos were then able to start accreting pebbles and planetesimals from their local surroundings, allowing them to grow. Once the planets reached $\sim$lunar masses, then they were able to migrate in the disc, encountering other planets, with some mutual interactions leading to collisions or ejections. Additionally, as the planets reached terrestrial--super-Earth masses, they were able to accrete gaseous envelopes. The planets mainly migrated inwards, towards the central cavity, where the buildup of material at the cavity acted as a migration trap, allowing planets to congregate there, again inducing collisions and ejections. The architectures of the final planetary systems, and the planets found within them, were similar to those formed in \citet{Coleman24}.

To highlight the evolution of a single simulated system, Fig. \ref{fig:sim_time} shows the temporal evolution of planet masses (top panel), and semimajor axes (bottom panel). We highlight those planets that are ejected from the simulation and with masses $m_{\rm p}>1\me$ with coloured lines, whilst the grey lines show all other planets. The crosses show the final planet masses and semimajor axes from when before they were ejected, with black crosses showing planets ejected through interactions with the binary stars, and the red crosses being through planet--planet interactions. Looking at the left hand side of Fig. \ref{fig:sim_time} it is clear that the planets are forming with masses $m_{\rm p}=10^{-3}$--$10^{-4}\me$, over the first 0.1Myr with those closest to the star forming first, around $\sim4\au$, with the outermost planet forming at$\sim 40\au$. Over the next 0.2 Myr, the planets accrete pebbles and planetesimals, whilst undergoing mutual collisions with other planets. This allowed a number of planets to reach terrestrial and super-Earth masses, where they could then begin to migrate through the disc. This evolution is nicely highlighted by the lower green and dark red lines in Fig. \ref{fig:sim_time} with the dark red line reaching a mass $m_{\rm p}\sim20\me$ after 0.2Myr through accreting large amounts of pebbles and planetesimals and colliding with multiple growing embryos.

With multiple super-Earths and Neptune mass planets forming, this leads to a number of interactions with a few planets being ejected. As well as low mass planets being ejected, three planets with masses $m_{\rm p}>1\me$ are ejected around 0.3 Myr. This cluster of ejections is due to the interactions between the green and dark red planets orbiting closer to central binary around $\sim1\au$. As these planets grew, they underwent runaway gas accretion, becoming giant planets. They then interacted with each other, forcing the green planet to move on to a more eccentric orbit, and to be scattered out to $\sim5$--$10\au$. Once there, it interacted with other planets, forcing them to be ejected from the system. This is highlighted by those planets ejected at around 0.3 Myr. Over the next 0.2 Myr, that giant planet migrated back in towards the central binary, with the inner giant being ejected after interacting with one of the stars. Additionally, as the green planet migrated inwards, it interacted with the dark blue planet, forcing it to interact with the binary and be ejected from the system. As the giant planet represented by the green line migrated closer to the binary, the planet shown by the light red line underwent runaway gas accretion, causing another episode of dynamical instabilities and ejections from the system. This resulted in the cluster of ejections at $\sim0.5$Myr. With fewer planets now orbiting in the system, the dynamical interactions were quenched and so the evolution of the system then followed a more orderly pathway, with the final system containing a giant planet orbiting at 7.8 $\au$, and a few sub-terrestrial and terrestrial mass planets orbiting between 22--117$\au$.

The ejections in the simulation described above were commonplace amongst the simulations presented in this work. Large frequencies of dynamical interactions, especially around the cavity close to the binary stars, acted to increase eccentricities and force planet to interact with more massive objects, e.g. giant planets or the binary stars themselves, and by ejected from the system. Additionally the example shows the abundance of ejections from circumbinary systems, whilst simultaneously showing the formation of the surviving circumbinary planets that have been seen in observations.

\subsection{Population statistics}
With Sect. \ref{sec:example_system} outlining an example of the formation and ejection of multiple planets from a circumbinary system, we now discuss the properties and the distributions of the population of ejected planets as a whole.
We begin by exploring the number of planets ejected over the course of each simulation.

\subsubsection{Frequency of ejected planets}
Recent work has estimated that the number of free-floating planets per star in the galaxy is $21^{+23}_{-13}$ with masses $m_{\rm p}>0.33\me$ \citep{Sumi23}. Previous works looking at a small range of parameters have also shown that between 3--9 planets are ejected from circumbinary systems around Kepler-16, Kepler-34 \citep{Coleman23}, and BEBOP-1 \citep{Standing23,Coleman24}.
The numbers of planets ejected per system here is in broad agreement with that expected from observations, and from previous theoretical work.
We find that each circumbinary system ejected $5^{+2}_{-3}$ planets with masses greater than 1$\me$ over the 10 Myr runtime of the system.

\begin{figure}
\centering
\includegraphics[scale=0.5]{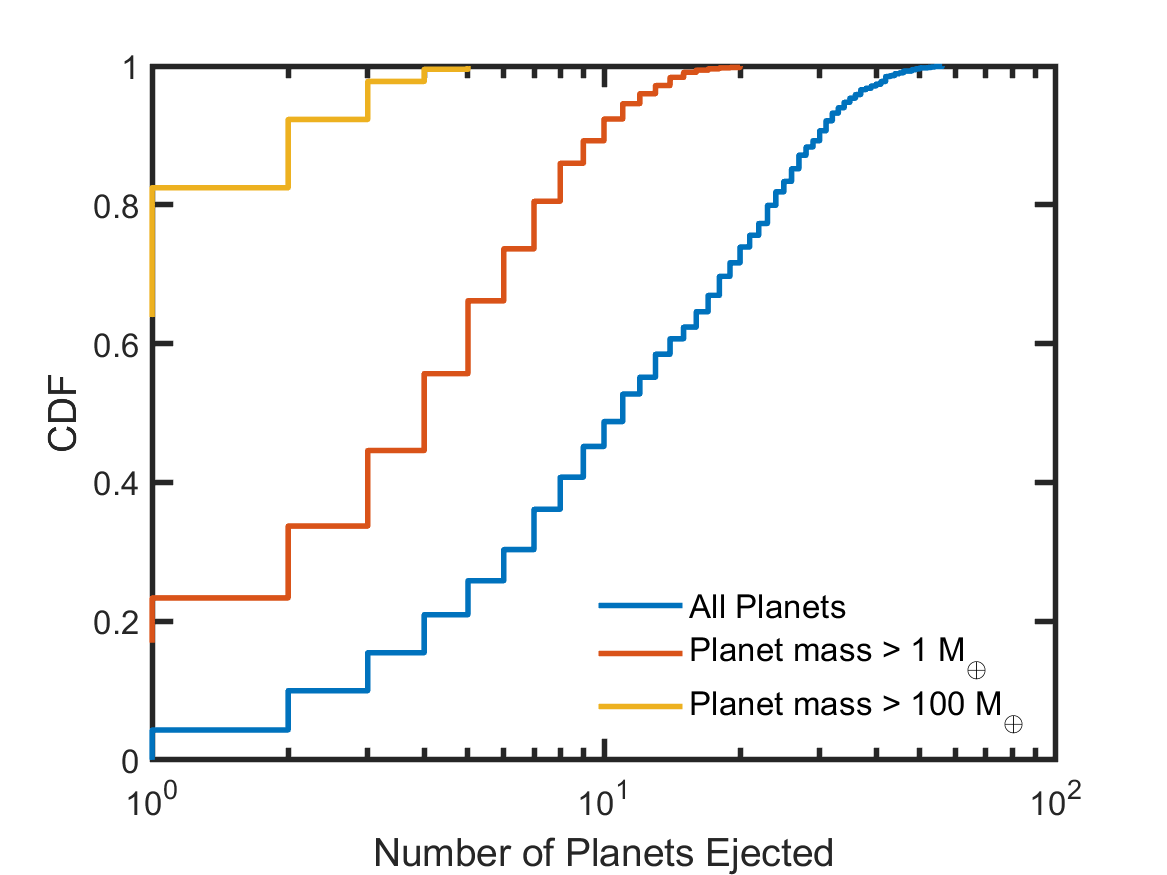}
\caption{Cumulative distribution functions for the number of planets ejected from the simulated circumbinary systems. The distributions are shown for all planets (blue line), for planets with masses $m_{\rm p}>1\me$ (red line), and for giant planets with masses $m_{\rm p}>100\me$ (yellow line).}
\label{fig:number_stats}
\end{figure}

In Fig. \ref{fig:number_stats} we show the cumulative distribution function for the number of planets ejected per system. The blue line shows the number of planets ejected per system for all planet masses, whilst the red and yellow lines show the number of planets ejected with masses $m_{\rm p}>1\me$ and 100 $\me$ respectively. As mentioned above the average number of planets ejected per system with masses $m_{\rm p}> 1\me$ is around 5 planets. These averages change to 14 planets of all masses per system, and 0.6 planets of masses $m_{\rm p}> 100\me$. In regards to the number of giant planets, this is consistent with the expectations from previous simulations around Kepler-16 and Kepler-34 \citep{Coleman23}. When comparing to observations, these values are towards the lower end on their predictions. However the predictions from \citet{Sumi23} is only based on few detections, hence the large uncertainty in their predictions, and so with more observed FFPs, by the Nancy Grace Roman Telescope \citep{Spergel15,Bennett18} or the Rubin Observatory \citep{LSST_2019} for example, then the observed predictions for the number of FFPs per star would become more robust.

Interestingly whilst the number of ejected planets may be less than that predicted from observations, the total mass in planets ejected per system is larger than in observations.
Based on FFPs found in microlensing surveys, \citet{Sumi23} predict a total mass of $80^{+73}_{-47} \me$. From our simulations here, whilst we only eject 5 planets on average that have masses $m_{\rm p}>1\me$, the average total mass of planets ejected per system is equal to $147^{+233}_{-84}\me$.
With 0.6 giant planets being ejected per system, they obviously contain the majority of the mass ejected per system. Their variance in masses also provides significant deviations of the total ejected mass.
Nonetheless, the simulations show that more mass is ejected from the systems compared to observations. These differences could arise form the lack of observed FFPs yielding poorly constrained estimates, but additionally the simulations presented here were based on a central stellar mass of $\sim 1.31 \msun$, whereas those planets observed in microlensing surveys could have originated around a wide variety of stars. Running such a population of circumbinary planet formation simulations following a stellar IMF \citep[e.g.][]{Kroupa01} is not yet feasible computationally, but simulating such a population could yield estimates that are more consistent with future observations.

\subsubsection{Temporal distribution of ejections}

\begin{figure}
\centering
\includegraphics[scale=0.5]{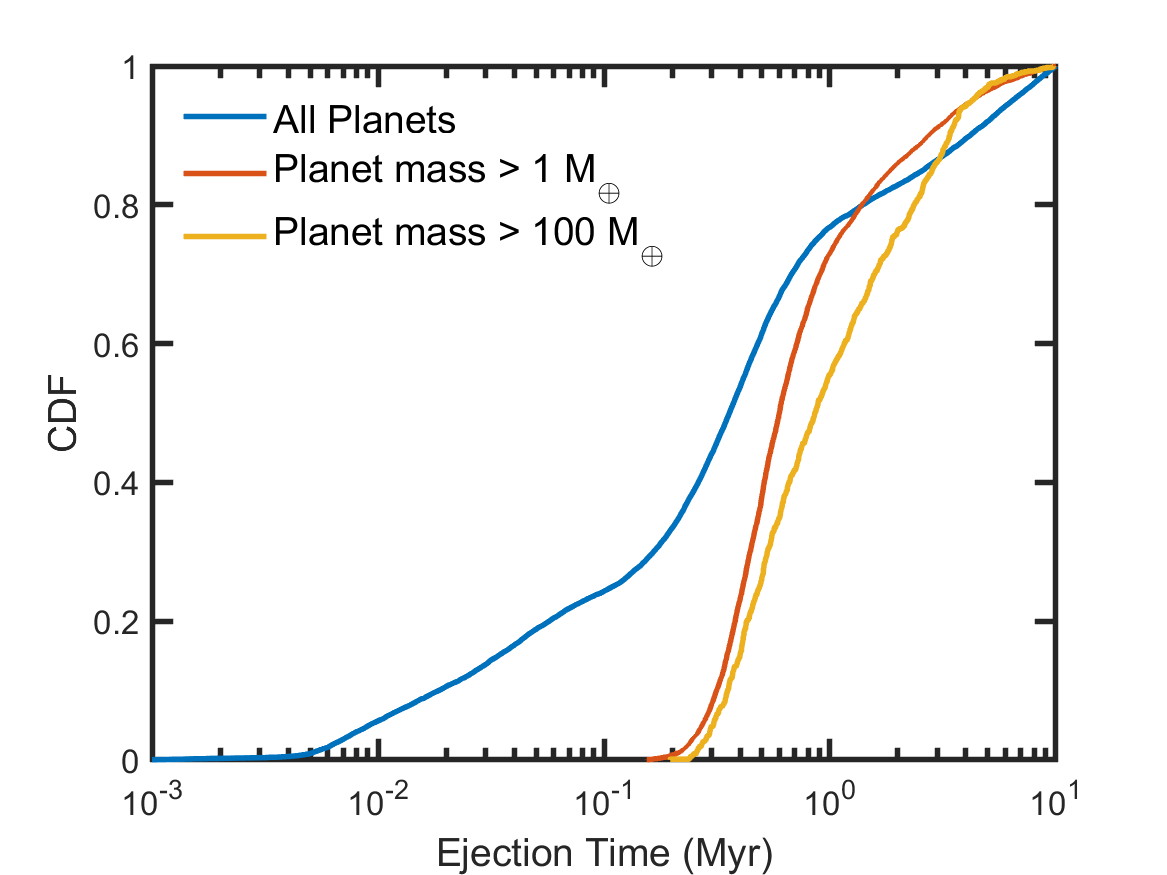}
\caption{Cumulative distribution functions for the time that planets are ejected from circumbinary systems. The distributions are shown for all planets (blue line), for planets with masses $m_{\rm p}>1\me$ (red line), and for giant planets with masses $m_{\rm p}>100\me$ (yellow line).}
\label{fig:EjectionTime}
\end{figure}

Recent observations of nearby star forming regions have found numerous planetary mass objects, down to a mass of $\sim$0.5 Jupiter masses, whilst their formation pathways are still an open question \citep{Pearson23}.
With planet and star formation still occurring in these regions, this makes the simulations presented here ideal to determine when planets are ejected. As discussed above, numerous planets are ejected over the 10 Myr simulation time, including the lifetime of the circumbinary discs. However, when the planets are ejected could be of use for observation surveys of nearby star forming regions to determine whether it would be expected to observe FFPs that originate in circumbinary systems.

To that end, Fig. \ref{fig:EjectionTime} shows the cumulative distribution function of planets as a function of their ejection time. The blue line shows the distributions for all planets, whilst the red and yellow lines again show the distributions for planets $m_{\rm p}>1 \me$ an $m_{\rm p}>100 \me$ respectively.
As can be seen from the red and yellow lines, it takes the systems at least $\sim 0.2$ Myr to begin to see these planets ejected from the systems. This is due to the need to form the initial planetary embryos, allow them to accrete pebbles and gas, and then interact with other planets, or the central binary itself, leading to their ejection.
This is not so much the case for lower mass planets that can be ejected much earlier in the disc lifetime, through interactions with the binary stars shortly after their formation.
Once sufficient planetary growth has occurred, it is clear that the rate at which planets are ejected remains roughly constant for the next 1--2 Myr. This arises due to the different growth time-scales of planet in different regions of the disc, as well as the dynamical nature of the ejected planet's evolution that causes it to move into a situation where may interact with other objects and be ejected from the system.
Therefore, Fig. \ref{fig:EjectionTime} shows that the majority of the ejections from circumbinary systems occurs in the first few million years of their lives, and so it would be expected to directly observe more FFPs in younger star forming regions with ages 0.5--3 Myr (e.g. ONC), than in older regions (e.g. Lupus).
Comparisons of the number of FFPs observed in these regions would provide valuable information on whether FFPs originate around circumbinary stars, or whether other processes, i.e. star formation processes, are responsible for these objects.
If star formation was delayed, or prolonged over a long period of time, then this would extend the times at which FFPs would be observable in star forming regions. Therefore, observations of large numbers of FFPs in older clusters, e.g. Upper Sco, would provide evidence for prolonged or multiple epochs of star formation in the region.

\subsubsection{Mass comparisons to bound planets}

\begin{figure}
\centering
\includegraphics[scale=0.5]{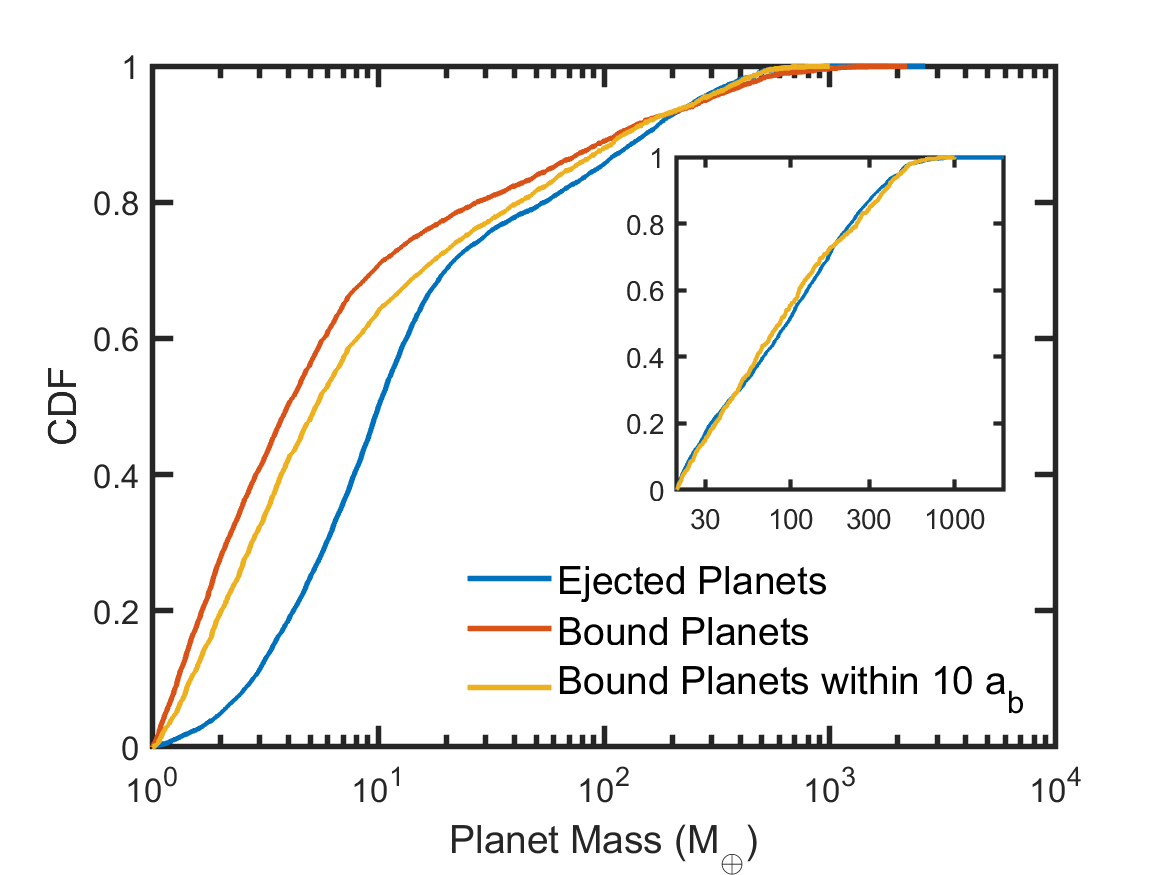}
\caption{Cumulative distribution functions for the mass of surviving planets with masses $m_{\rm p}>1\me$. The distributions are shown for all planets ejected from the systems (blue line), for all remaining bound planets (red line), and for all bound planets with semi-major axes $a_{\rm p}\le 10a_{\rm bin}$. The inset plot shows the cumulative distribution function for planets with masses $m_{\rm p}>1\me$, highlighting the similarity between ejected planets and bound planets with semi-major axes $a_{\rm p}\le 10a_{\rm bin}$.}
\label{fig:mass_comparison}
\end{figure}

Whilst direct observations of FFPs in the Trapezium Cluster have only found planets down to a mass of 0.5 Jupiter masses \citep{Pearson23}, microlensing surveys are now finding planets with super-Earth or even terrestrial masses \citep{Mroz19,Mroz20,Koshimoto23}.
With only a handful of sub-Jupiter mass FFPs found though, a robust estimation of the underlying mass distribution is not yet forthcoming, though future observations with the Roman Telescope or Rubin Observatory will bridge these uncertainties and deliver a mass distribution of FFPs. To that end, it is useful to understand the mass distribution of planets ejected from binary systems, and more importantly, how that distribution relates to the distribution of the remaining planets in the systems that can be detected through other detection methods (e.g. transit or radial velocity surveys). Understanding any differences in these populations will aid in the modelling of planet formation and evolution.

Figure \ref{fig:mass_comparison} shows the cumulative distribution function of planet mass for those planets with masses $m_{\rm p}>1\me$.
The blue line in Fig. \ref{fig:mass_comparison} shows the distribution for planets ejected from the systems, whilst the red and yellow lines show the distributions for all bound planets, and for bound planets with semi-major axes $a_{\rm p}\le10\times a_{\rm bin}$, i.e. with orbital periods amenable for detection by transit or radial velocity surveys.
It is clear to see that for all three distributions the bulk of the planets that remain bound and ejected are of low mass with 70--80\% of planets having masses $m_{\rm p}\le20\me$. There are differences there however, where there are few terrestrial planets ejected, and considerably more super-Earths and Neptune mass planets than for the distribution of the bound planets. For the bound planets the increase in the distributions are steady from terrestrial mass up to $\sim 15\me$ showing where pebble accretion dominates the formation of these planets. However, for the ejected planets, more super-Earth and Neptune mass planets are ejected since as multiple planets of these masses form in a single system, they can excite eccentricities and force other planets on to orbits where they interact with the binary stars and get ejected. Systems of Earth-mass planets are less likely to undergo this evolution since they need to be orbiting with significantly closer proximity in order to dynamically excite each other on to the required eccentric orbits for interactions with the binary stars.

Interestingly for planet masses greater than 20 $\me$, there is very little difference between the mass distributions of ejected planets (blue line) and those that remain bound to the star but orbit within $10\times a_{\rm bin}$ (yellow line). This shows that of the planets that form in those systems, similar fractions of planets as a function of planet mass remain bound and are ejected.
The inset plot in fig. \ref{fig:mass_comparison} shows the cumulative distribution function for planets with masses $m_{\rm p}\ge20\me$ for ejected planets and those planets that remain bound but with semi-major axes $a_{\rm p}\le10\times a_{\rm bin}$. It clearly shows the similarity between the two distributions for plants with masses greater than 20$\me$.
This result indicates that the observed mass distributions of FFPs through microlensing surveys should be similar to the mass distribution of planets seen in transit and radial velocity surveys. Obviously this would require a large number of  observed planets, since the stellar populations that account for the observed planets would also need to be consistent with each other. Additional effects that affect the formation such as the turbulence in the disc, and the local star forming environment, which we will discuss later, can also affect the distributions of planets and so differences of the mass distributions in these planets could also indicate different initial conditions for such planets. An example of this could include comparing FFPs that formed and are found in the galactic bulge, to those planets that formed in nearby low-mass star forming regions. Ultimately though, the observed mass distributions of these Neptune to Jupiter mass planets can inform significantly on the formation history of such planets.

\begin{figure}
\centering
\includegraphics[scale=0.5]{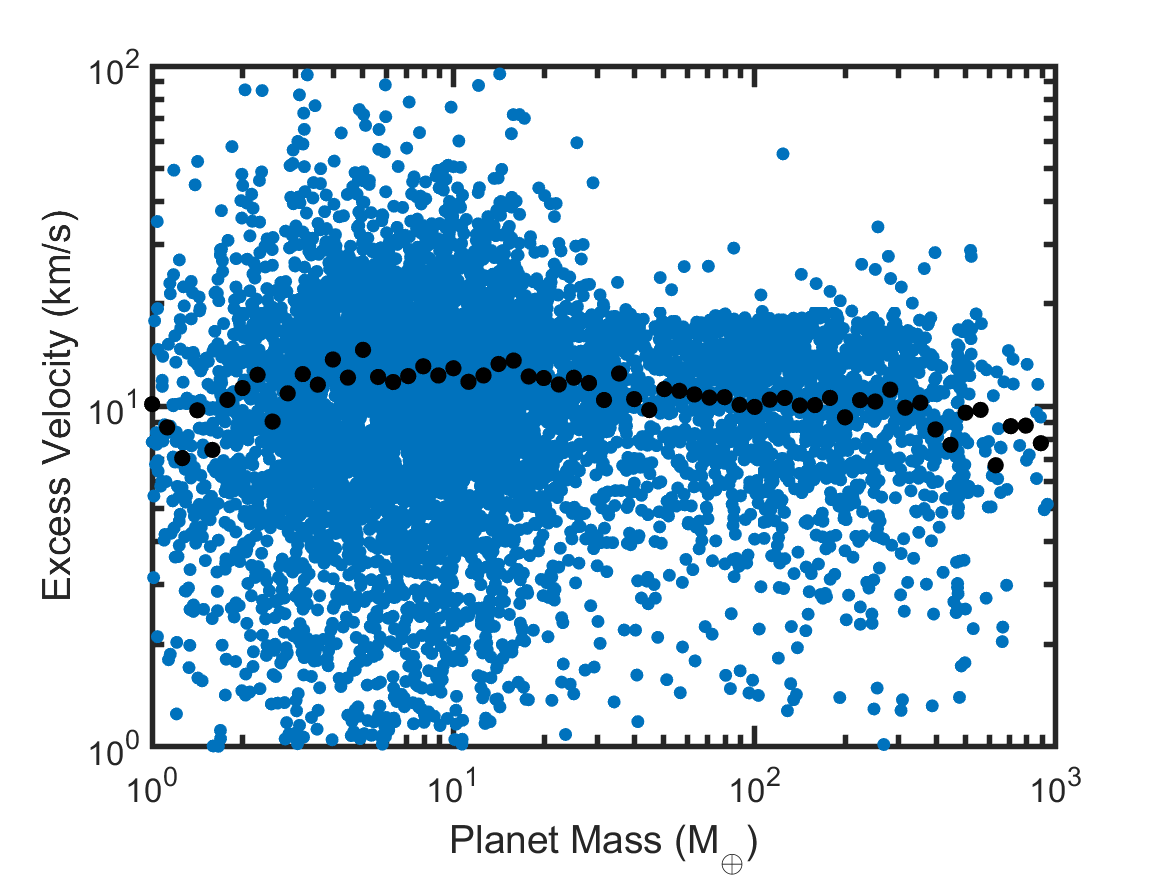}
\caption{Excess velocities of ejected planets as a function of planet mass. Blue points show individual planets, whilst black points show the average excess velocity within mass bins covering 0.05 dex.}
\label{fig:mass_excess_velocity}
\end{figure}

\subsubsection{Velocity signatures}

The mass distributions shown above are useful for comparing with FFPs found with microlensing surveys. However they are not as useful to compare with the more massive planets found with direct imaging by for example JWST or Roman. One property that is useful to compare with such observations is the observed velocity differences to other stars in the star forming regions. Numerous works have explored the velocity dispersions of stars in nearby star forming regions. For example the velocity dispersions found in the Orion Nebula Cluster is equal to $\sigma_v\sim$1--3 $\rm kms^{-1}$ \citep{VanAltena88,Kim19,Theissen23}. It would be expected that planetary mass objects that form at the tail end of the star formation process would have similar velocity dispersions. However if such objects formed through planet formation processes and were ejected from their systems by other planets or a binary star for example, then their velocity dispersions could be considerably different.

For an object to be ejected from a system, their velocity must be larger than the escape velocity. If such objects are launched on these hyperbolic orbits with more velocity than the escape velocity, then they will have an excess velocity which can be measured as these objects move through the local regions. To calculate the excess velocity we use the hyperbolic orbital elements of the planet as they are removed from the simulations when they reach 1000$\au$. More specifically it is equal to:
\begin{equation}
v_{\infty} = \sqrt{\dfrac{-v^2}{a(2/r-1/a)}}
\end{equation}
where $v$, $r$, and $a$ are the planet's velocity, distance from the barycentre, and semi-major axis when removed from the simulation.

In Fig. \ref{fig:mass_excess_velocity} we show the planet excess velocities as a function of the ejected planet mass. The blue points represent individual planets, whilst the black points show a binned average with a bin size equal to 0.05 dex.
As can be seen in Fig. \ref{fig:mass_excess_velocity} there is a large variance in the excess velocity of objects, especially for those of lower mass. For lower mass planets, they can be deflected on to extremely close encounter orbits with one of the binary stars, which can result in extremely fast ejection velocities. This gives rise to the excess velocities for these objects reaching $\sim $100 $\rm kms^{-1}$. Looking at the black points showing the average values, it can be seen that the average ejection velocities increases from 8 $\rm kms^{-1}$ for Earth mass planets up to 13 $\rm kms^{-1}$ for 15 $\me$ planets. This increase is due to multiple planets forming and mutually interacting close to the edge of the central cavity, allowing planets to strongly interact with the binary stars. Interestingly the variance in the excess velocity, and the average velocity decreases as the ejected planet mass increases.
By looking at the blue points it is clear that the majority of planets with masses $m_{\rm p}>30\me$ are ejected from their systems with $v_{\infty}\le $20 ${\rm kms^{-1}}$, whilst the average excess velocity falls from 12 $\rm kms^{-1}$ for 30 $\me$ planets down to 8 $\rm kms^{-1}$ for 500 $\me$ planets.
When planets reach these masses, the damping forces they feel from the circumbinary discs act to reduce their eccentricities resulting in weaker interactions with the binary, whilst there are also fewer planets of a similar mass that can force them to interact with the binary through strong dynamical encounters. Nonetheless, it is interesting here that the average velocity of planets ejected form circumbinary systems is around 10 $\rm kms^{-1}$, a factor few times larger than the velocity of other objects formed in the star forming regions.

Whilst Fig. \ref{fig:mass_excess_velocity} showed the excess velocity and masses of ejected planets, it is not able to answer the question if there are differences between those values for planets ejected through interactions with the binary stars or through interactions with planets further out in the system. The origins of the ejection event are important, since those that are ejected through interactions with other planets, far from the binary stars, will have properties akin to planets ejected in single stellar systems of a similar central mass. Dynamically, when not close to the binary, the escape velocity from mutual planetary interactions has to be larger than the escape velocity of the orbit, identical to the requirements in single stellar systems. Therefore it is possible to compare the properties of ejected planets from interactions with the binary, i.e. binary induced, to those where the binary is not involved akin to single stellar systems. In order to make this comparison we must first differentiate within our ejected sample those planets that are ejected through interactions with the binary, and those from further out in the disc. Using the hyperbolic orbital elements, we can calculate the pericentre of the orbit. Under the assumption that the last significant gravitational interaction was what led to the ejection of the planet, then the pericentre of the orbit is the approximate location of this event, thus informing us where the planet was ejected from.

\begin{figure}
\centering
\includegraphics[scale=0.5]{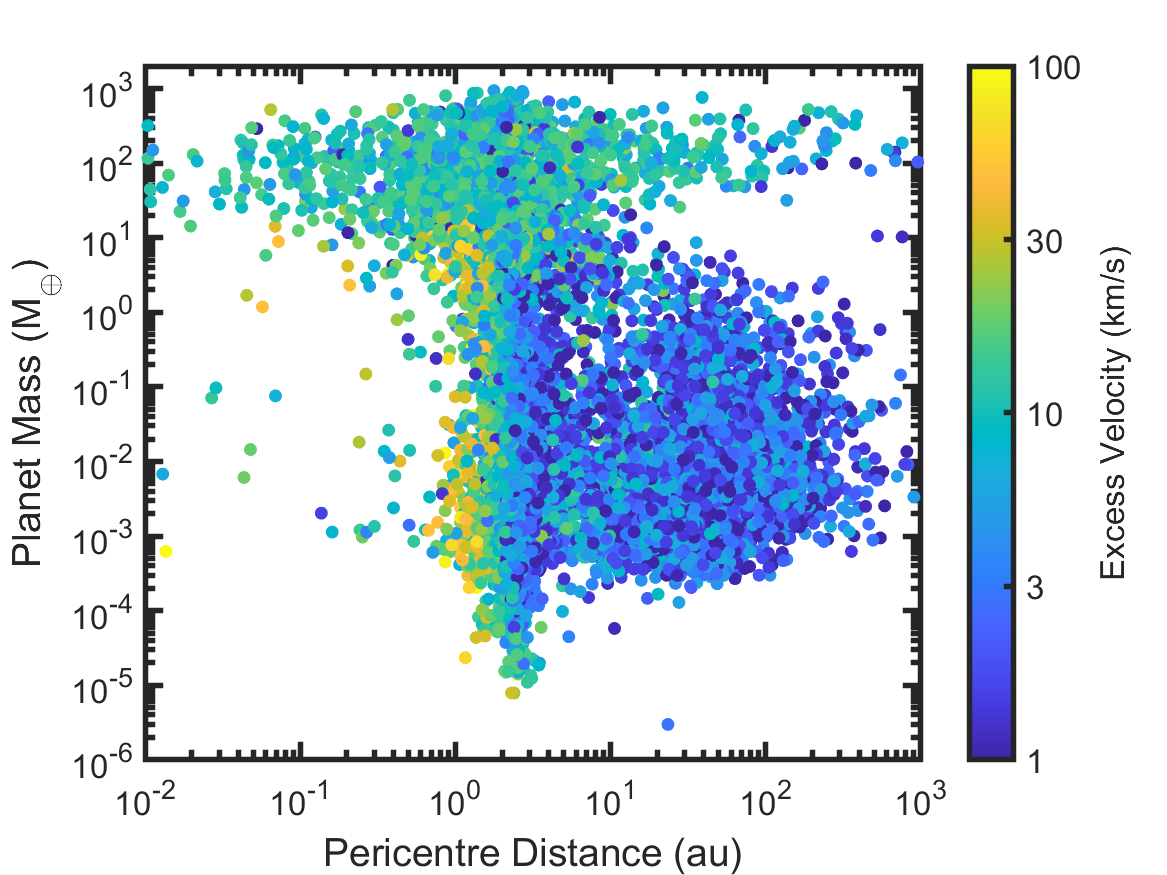}
\caption{Planet mass versus the pericentre distance for ejected planets. The pericentre distance gives an approximate location for the final interaction that led to the ejection of the planet. The colour coding shows the excess velocities of the planets after they have been ejected.}
\label{fig:EjectionLocations}
\end{figure}

Figure \ref{fig:EjectionLocations} shows the ejected planet mass versus the pericentre distance, i.e. the ejection location. The colours show the excess velocity of the planets. From Fig. \ref{fig:EjectionLocations} we can immediately identify three separate populations.
The first population of planets are those with pericentre distances $p_{\rm p}>10\au$, and masses, $10^{-4}\me<m_{\rm p}<10\me$. These low mass planets and planetary embryos are ejected far from the binary stars, typically through mutual interactions with more massive objects, i.e. Neptune to Jupiter mass planets depending on their orbital separation. These are the planets that will be most like those that are ejected in single stellar systems, since the binary played a negligible role in the ejection of these objects. Interestingly the planets ejected in this region, typically have excess velocities around 1--10 $\rm kms^{-1}$, showing that they are weakly ejected from the system, and would have similar velocity dispersions to surrounding objects in their local region.

The second population can be found closer to the central binary and encompasses the low mass planets with pericentre distances $p_{\rm p}<20 a_{\rm bin}$, and masses, $10^{-5}\me<m_{\rm p}<10\me$. Previous works have shown that circumbinary discs are dominated through interactions with the central binary within $20a_{\rm bin}$, resulting in the formation of a cavity and a trapping location for migrating planetesimals, planetary embryos, and planets \citep{Coleman22b,Coleman23}. This allows objects to concentrate in this region before being excited on to orbits that interact with the central binary, and thus then ejected. The effect of the binary is also extremely clear here on the excess velocity, with those objects that are ejected closer to the cavity edge, located typically around 1$\au$ being ejected at much larger velocities, shown by the orange and yellow points in Fig. \ref{fig:EjectionLocations}. Such ejections would not be possible in single stellar systems since the objects are too deep in the gravitational wells to be ejected by most planets.

The third and final population is those planets with masses $m_{\rm p}>10\me$. These planets have a wide range of pericentre distances from $0.01\au$--1000$\au$. They are also typically ejected with excess velocities between 3--30 $\rm kms^{-1}$. These ejection locations and velocities are very different to the two other populations, especially the pericentre distances. This arises due to numerous complications in calculating the ejection location. Firstly, the damping that acts on these planets can result in subtle changes to the planet's velocity and thus affect the excess velocity derived from the planet's orbital elements when it reaches 1000$\au$. The other main complication arises through the fact that these simulations involve multiple planets, and typically when planets are excited on to eccentric orbits, they are done so by sufficiently massive bodies. As planets are placed on to hyperbolic, escaping, orbits by the binary stars, there is a non-negligible possibility that these planets interact with other massive objects as they exit the inner parts of the system. Such interactions can again act to speed up or slow down the ejecting planet, sometimes causing it to be bound to the system again. Nevertheless these interactions add significant contamination to determining where the planets are ejected from. Aside from those complications, it is still important to acknowledge that these more massive planets have significantly large excess velocities, much greater than what is see for stellar populations in star forming regions. Such large differences in velocity dispersions of FFPs if shown would indicate that they originated in circumbinary systems.

\begin{figure}
\centering
\includegraphics[scale=0.5]{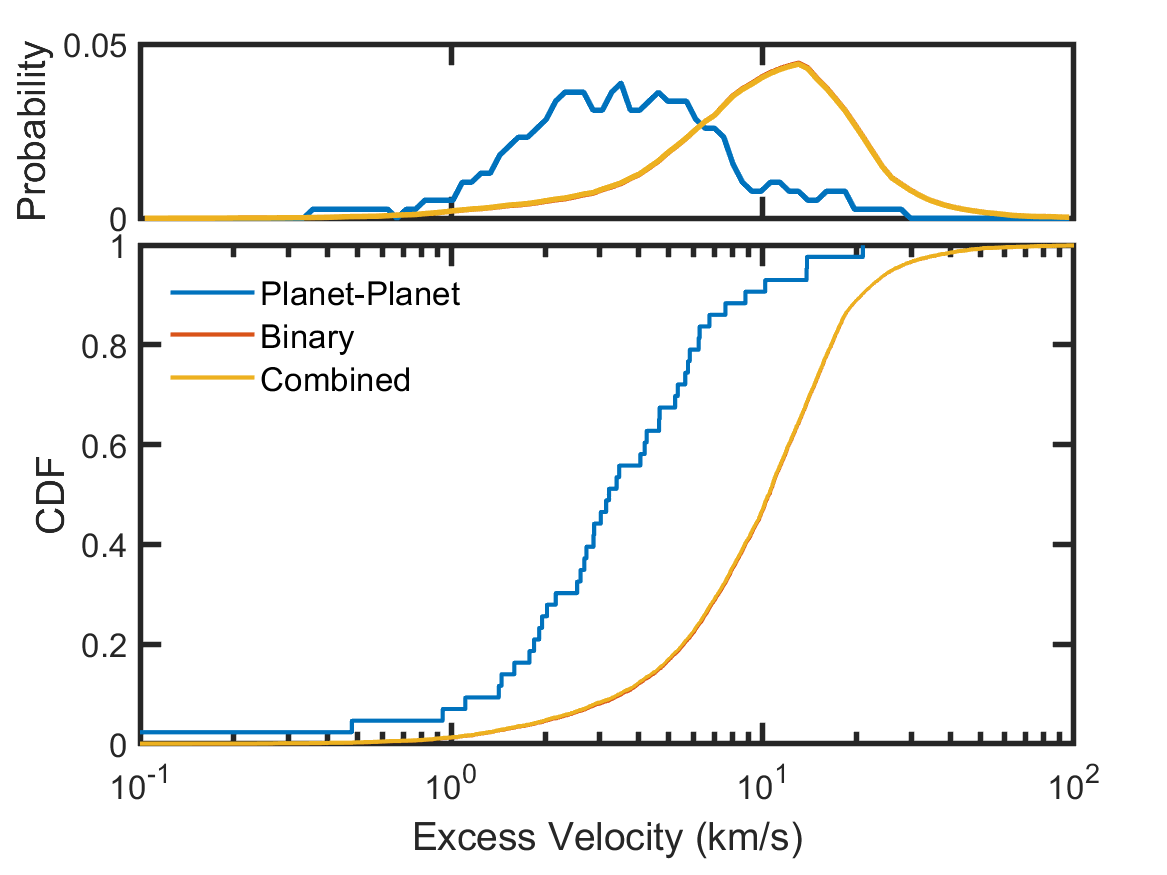}
\caption{Cumulative distribution functions (bottom panel) and probability distribution functions (top panel) of planet excess velocities for ejected planets with masses $m_{\rm p}>1\me$. The lines differentiate between planets ejected through planet--planet interactions (blue line), planets ejected through interactions with the binary stars (red line), and a combined distribution (yellow line).}
\label{fig:excess_velocity_stats}
\end{figure}

To further the question of determining if it is possible to tell if FFPs originate in circumbinary systems, we plot the cumulative distribution function of the ejected planets excess velocities in Fig. \ref{fig:excess_velocity_stats}. We only include planets with masses $m_{\rm p}>1\me$ to be more consistent with what may be observable with future observations. We show the CDF for all planets with the yellow line, whilst we split the population into two groups, those that are scattered through planet-planet interactions similar to the first population described above but including more massive planets (blue line), and those that are ejected through interactions with the binary stars (red line). For the planets that have complicated histories, i.e. those that may have interacted with other objects on their escape trajectory, we check whether their semi-major axis within the last 10,000 years of their time bound in the simulations was situated within $10a_{\rm bin}$ of the central binary.
It is clear from Fig. \ref{fig:excess_velocity_stats} that there is a large difference in the velocity dispersions between those planets ejected through planet--planet interactions and those through interactions with the binary itself. Indeed, for planet--planet interactions, the average velocity that planets were ejected at was equal to $4.54^{+1.21}_{-2.58}\rm kms^{-1}$, with the limits showing the interquartile range. This is in stark contrast to those ejected through interactions with the binary where the average velocity was equal to $11.97^{+3.37}_{-5.54}\rm kms^{-1}$. Figure \ref{fig:excess_velocity_stats} therefore shows that the planets that are ejected through interactions with the binary stars have velocities roughly three times larger than those through planet--planet interactions, a result which should also hold when comparing ejections from binary systems to single stellar systems. With such large velocities, it would also be expected that these planets would have insufficient time to standardise their velocities to that of the stars in the star forming region. This is due to the ejection times of planets in most star forming regions (e.g. 0.5 Myr for a region of width $\sim$few parsec) being much shorter than the relaxation time \citep{Wang15}.
Additionally, with such a large velocity signature, compared to the background stars in the local region, finding such a large velocity dispersion would be a clear signature of circumbinary planet formation, and ejection of FFPs.

The final point to take from Fig. \ref{fig:excess_velocity_stats} is the frequency at which ejections occur for both populations. As can be seen there is very little difference in the velocity dispersions between the binary and combined lines in Fig. \ref{fig:excess_velocity_stats}. This shows that the binary induced ejections are contributing the most to the observed velocity dispersions. Indeed, when looking at the absolute numbers of ejections, binary induced ejections contribute 99.4\% of all ejections from the system, highlighting how difficult and rare it is for planet--planet interactions to lead to ejections of an object.
From observational surveys, 5--10$\%$ of Solar type stars in binary systems are of configurations comparable to those explored here, i.e. close-binaries \citep{Offner23}. Combining this with Solar-type stars being found to contain 0.6 companions per stars \citep{Offner23}, then for every close binary system explored here, there would be 7--14 singular Solar type stars.
Coupling this with the numbers of ejected planets per system described previously, and the predicted numbers of FFPs per star based on microlensing surveys, this would again indicate that binary systems, instead of single stellar systems, are responsible for the formation of FFPs.

\begin{figure*}
\centering
\includegraphics[scale=0.5]{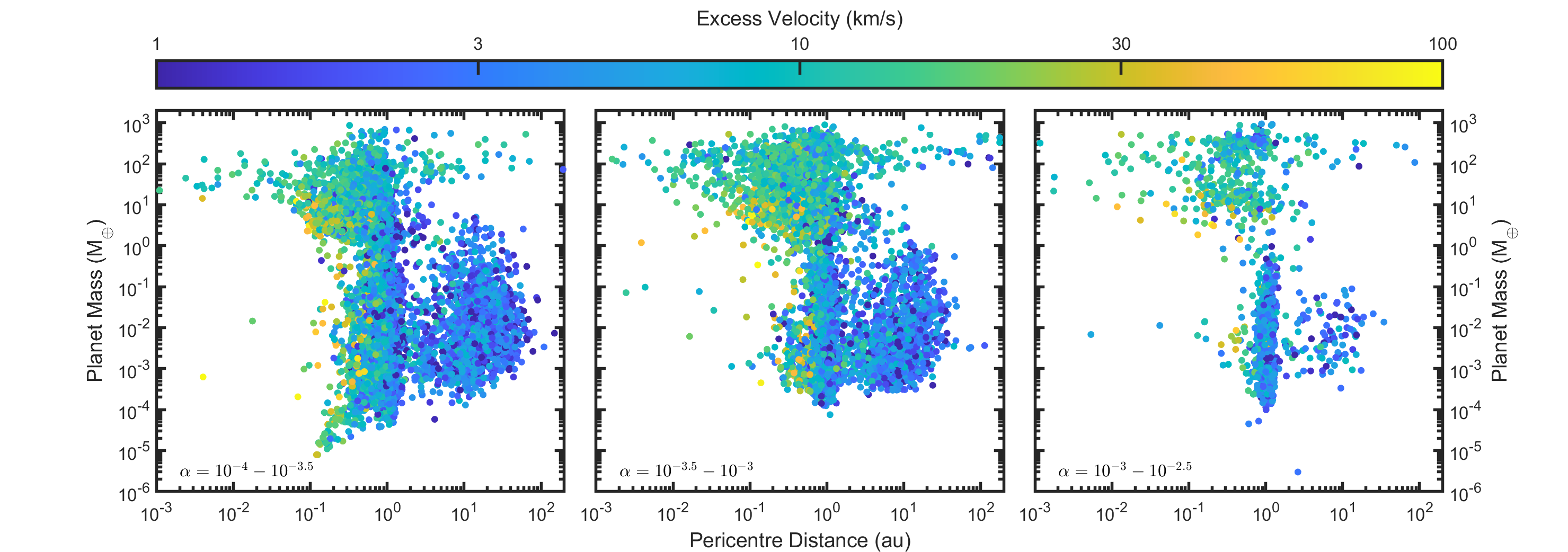}
\caption{Planet mass versus pericentre distance (ejection location) for all planets formed in simulations with viscosity $\alpha$ values between: $10^{-4}$--$10^{-3.5}$ (left-hand panel), $10^{-3.5}$--$10^{-3}$ (middle panel), and $10^{-3}$--$10^{-2.5}$ (right-hand panel). The colour coding denotes the excess velocities of the planets after they have been ejected.}
\label{fig:alpha_triple}
\end{figure*}

Additionally, stellar flybys could be an efficient production mechanism of FFPs \citep{Wang24}. However, the frequency of stellar flybys is determined by the number density of stars in a star forming region. For nearby star forming regions, e.g. ONC, the number density reaches $10^4 \rm pc^{-3}$ \citep{Winter18}, which equates to a background flux of $10^4 \rm G_0$ in this work. For this number density, the closest encounter for stellar flybys lies between 100--1000$\au$, much too far from the inner system to drive efficient ejections of planets \citep{Winter18}. For less dense clusters, the closest approaches are even larger, further reducing the effectiveness of stellar flybys as FFP production mechanisms.

Confirming some of the other observational signatures, such as the velocity dispersion would further add confidence to the conclusion that binary stars are essentially dominant FFP factories.

\section{How do simulation parameters affect the distributions}
\label{sec:parameter_effects}
Section \ref{sec:results} explored the population statistics as a whole that arose from the simulations. We now determine the effects that different initial parameters have on the observable distributions of FFPs. These include the level of turbulence in the disc and the external photoevaporation rate, which have previously been found to significantly affect the evolution of protoplanetary discs and the planets that form within them \citep{Coleman22,Coleman23}. We will also explore the effects of the binary separation on the distributions.

\subsection{How turbulence affects the FFP population}
The level of turbulence in the disc is determined by the strength of $\alpha$ where in our simulations we have used values between $10^{-4}$--$10^{-2.5}$, consistent with observations \citep{Rosotti23}.
In Fig. \ref{fig:alpha_triple} we show three separate plots similar to Fig. \ref{fig:EjectionLocations} where each plot shows the planet mass versus the pericentre distance (ejection location), with the colours showing the excess velocity of the ejected planets. The left-hand panel shows for planets that formed in discs with $10^{-4}\le\alpha<10^{-3.5}$, the middle panel being $10^{-3.5}\le\alpha<10^{-3}$ and the right-hand panel being for planets in discs with $10^{-3}\le\alpha<10^{-2.5}$. The different panels essentially show the effects of increasing turbulence from left to right.

\begin{figure}
\centering
\includegraphics[scale=0.5]{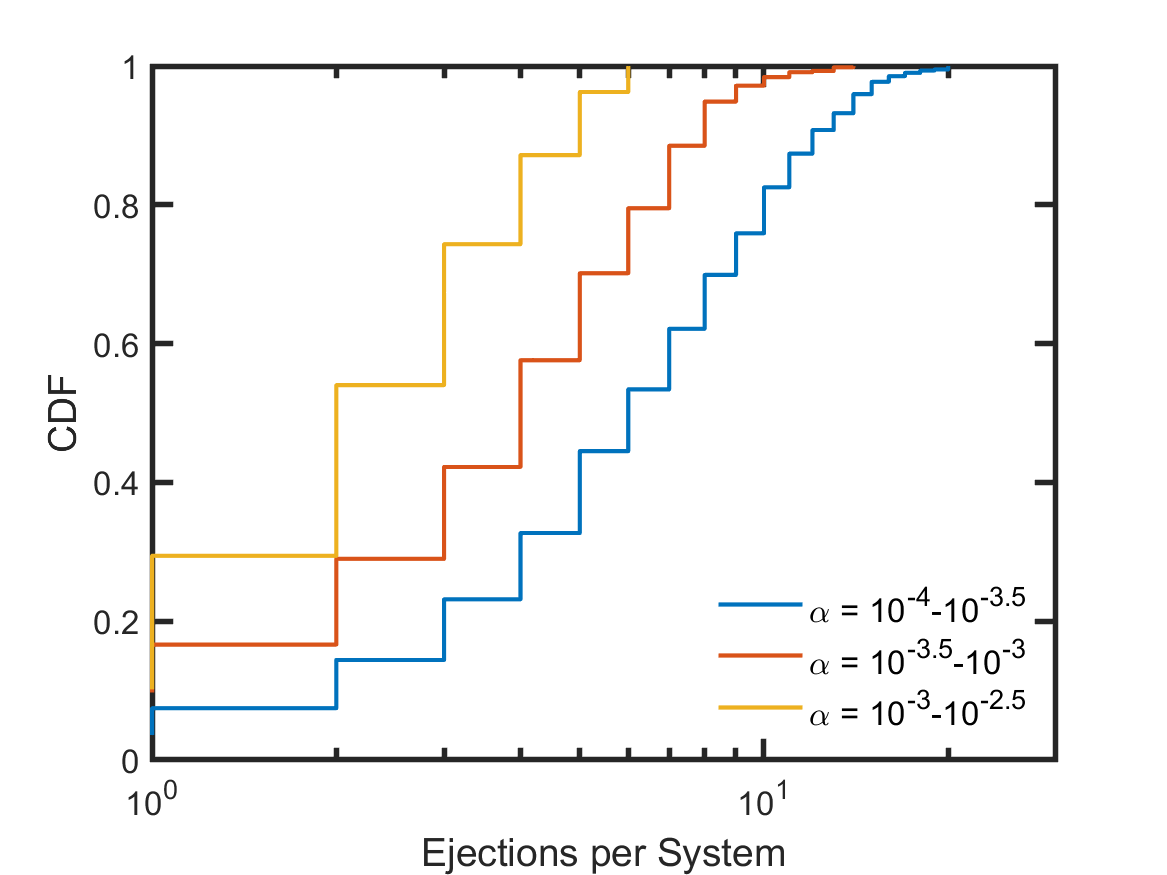}
\caption{Cumulative distribution functions for number of planets ejected from each system. The level of turbulence in the disc is denoted by the colours with the blue line showing discs with $10^{-4}\le\alpha<10^{-3.5}$, the red line showing discs with $10^{-3.5}\le\alpha<10^{-3}$, and the yellow line being for discs with $10^{-3}\le\alpha<10^{-2.5}$.}
\label{fig:alpha_ejections}
\end{figure}

Comparing the panels in Fig. \ref{fig:alpha_triple} it is clear that there are differences depending on the level of turbulence. The main differences arise in the outer regions of the disc where there are fewer planets ejected through planet-planet interactions as $\alpha$ increases (including sub-terrestrial mass planets). Additionally, fewer planets are ejected through interactions with the central binary. The main causes for these reductions is on the effectiveness of forming the initial planetary embryos and planetesimals decreasing as $\alpha$ increases. As $\alpha$ increases, the turbulence in the disc stirs pebbles to higher altitudes, resulting in a less dense pebble layer in the midplane. Since one of the criterion for the gravitational collapse of a pebble cloud to occur is that the local solids-to-gas ratio must exceed unity, the reduction in density due to higher $\alpha$ values makes this harder to achieve. Thus fewer planetesimals and planetary embryos are able to form and grow, and so there are then fewer planets interacting with each other or the binary and ejected from the system.

Figure \ref{fig:alpha_ejections} shows the number of all planets that are ejected from discs with different levels of turbulence, with blue lines showing low $\alpha$ values, and red and yellow lines showing intermediate and high values respectively. The reduction in the number of planets ejected as $\alpha$ increases is clear to see, with discs evolving with low $\alpha$ values ejecting $\sim$ 3 times more planets on average than those with large $\alpha$ values. These trends show that if observations can accurately constrain the number of FFPs per star, then they can also provide insights into the fundamental properties of protoplanetary discs.

\begin{figure}
\centering
\includegraphics[scale=0.5]{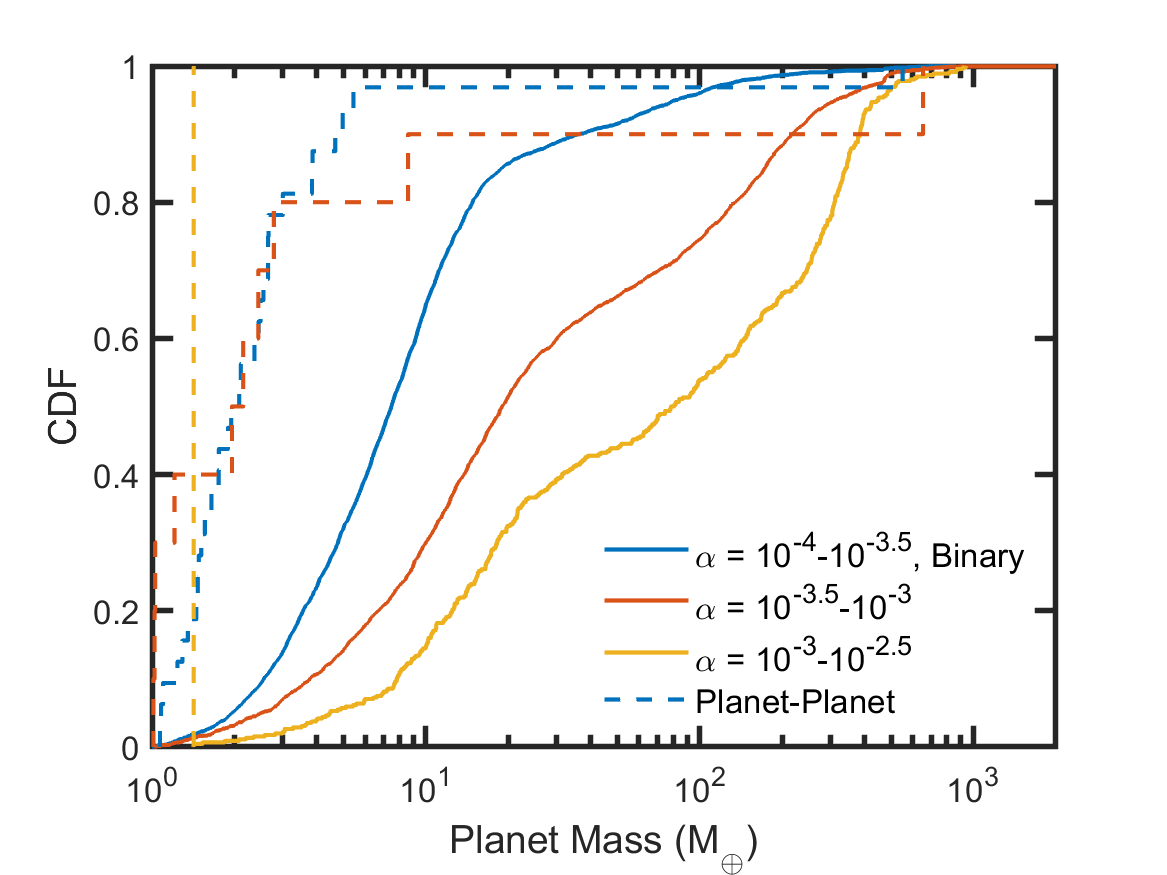}
\caption{Cumulative distribution functions for the mass of surviving planets with masses $m_{\rm p}>1\me$. The distributions are shown for all planets ejected from the systems with $10^{-4}\le\alpha<10^{-3.5}$ (blue line), $10^{-4}\le\alpha<10^{-3.5}$ (red line), and $10^{-4}\le\alpha<10^{-3.5}$ (yellow line). Solid lines show planets ejected from the vicinity of the binary, whilst dashed lines show planets ejected through planet--planet interactions.}
\label{fig:alpha_mass}
\end{figure}

\begin{figure*}
\centering
\includegraphics[scale=0.5]{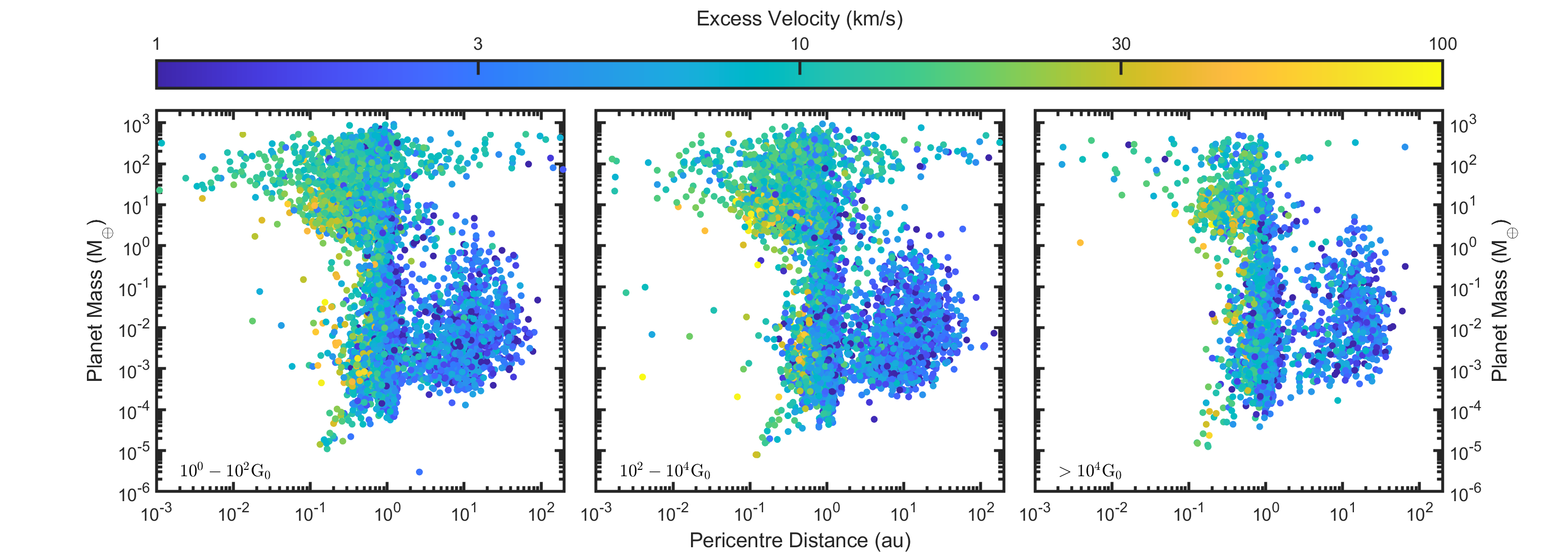}
\caption{Same as Fig. \ref{fig:alpha_triple}, but the panels show for planets ejected in weak UV environments ($<10^2 \rm {G_0}$, left-hand panel), intermediate UV environments ($10^2$--$10^4 \rm {G_0}$, middle panel), and strong UV environments ($>10^4 \rm {G_0}$, right-hand panel).}
\label{fig:environment_triple}
\end{figure*}

As figs. \ref{fig:alpha_triple} and \ref{fig:alpha_ejections} showed that fewer planets were ejected from systems as the viscosity parameter $\alpha$ increases, this also has an effect on the mass distributions of planets that are ejected.
Figure \ref{fig:alpha_mass} shows the mass distributions for planets above an Earth mass ejected from discs with different strengths of $\alpha$, as shown by the different coloured lines.
The solid lines show those planets ejected through interactions with the central binary, whilst the dashed lines show those planets lost through planet-planet interactions. An interesting feature that arises from Fig. \ref{fig:alpha_mass}, especially for the binary induced ejections, is that as the strength of turbulence $\alpha$ increases, the mass distribution of ejected planets shifts to more massive planets. Indeed in discs with strong levels of turbulence, $\alpha\ge 10^{-3}$, 46 \% of planets ejected have masses $m_{\rm }>100\me$, whereas in discs with weak levels of turbulence, giant planets make up 4\% of those ejected.
Therefore more lower mass planets are ejected in weaker turbulent discs. The main cause of this change in distributions as a function of $\alpha$ is due to planets being able to more easily form in low turbulent discs, as the pebbles are more confined to midplane of the disc, aiding both planetesimal/embryo formation, as well as pebble accretion rates. As planets reach the pebble isolation mass, their accretion rate slows since gas accretion is inefficient there. This results in numerous terrestrial and super-Earth mass planets occupying similar areas of space, mutually interacting and driving up eccentricities. They then interact with the binary and are ejected from the system. In discs with higher $\alpha$, this is not so much the case, since there are fewer planets that initially form and grow, and so a greater percentage of planets can undergo runaway gas accretion, becoming giant planets, before they are ejected from the systems.

Interestingly, the distributions in Fig. \ref{fig:alpha_mass} appears to become more bi-modal as $\alpha$ increases, with super-Earth-Neptune mass and giant planets making up the majority of planets ejected in high $\alpha$ discs. The dearth of planets with masses between $\sim30 \me$ and $\sim 100\me$ is a signature of runaway gas accretion, since one the mass of the gaseous envelope becomes comparable to the core mass, the self gravity of the envelope allows it to quickly contract, driving up accretion rates \citep{CPN17}. This runaway gas accretion halts when the supply of gas to the envelope becomes limited since the planet opens a gap in the disc. However in low $\alpha$ discs, since the majority of planets ejected are only of super-Earth to Neptune mass, since there is greater frequencies of dynamical interactions, the signature of runaway gas accretion is effectively diminished. Additionally, with low values of $\alpha$, planets are more easily able to open gaps in the discs. Observing such a bimodality in observations would therefore then give hints as to the underlying properties of protoplanetary discs.

\subsection{What is the role of the local stellar environment}

Previous studies have shown that the local environment influences how planets form \citep{Winter22,Qiao23} and therefore also on the resulting planet populations \citep{Standing23,Coleman24}. Those works found that as the strength of the local environment increased, the possibility for more massive planets to form and grow diminished. This was due to the reduction in accretable solids since external photoevaporation from the nearby massive stars effectively truncated the disc to small sizes. Within our simulations presented here, we find similar effects on the planets that form, and thus on the final planetary systems. There are also noticeable effects on the properties of free floating planets that escape from systems in different environments, of which we will now discuss.

Similar to Fig. \ref{fig:alpha_triple}, Fig. \ref{fig:environment_triple} shows three separate plots of ejected planet mass versus the pericentre distance (ejection location) with the colours showing the remaining excess velocity. Instead of the different panels showing the strength of turbulence in the disc, this time they show planets that formed in different environments. Again, this increases from left-to-right with the left-hand panel showing planets that formed in weak UV environments $(<10^2 \rm G_0)$, the middle panel showing for intermediate environments $(10^2 {\rm G_0}$--$10^4 {\rm G_0})$, and the right-hand panel being for planets forming in strong UV environments $(>10^4 \rm G_0)$. When moving form left to right on the plot, it is clear that the left-hand and middle panels are similar in terms of the populations of planets at different ejection locations and of specific planet masses, but there are significant differences for the right-hand panel, the strongest UV environments. Most noticeably is the reduction in giant planets ejected at the top of the panel, with the majority of giant planets this time being ejected from around the edges of the circumbinary cavities. Both, the reduction of giant planets ejected, and the confinement of the ejection location, is a result of the reduction in solids available for accretion. With less solids, fewer giant planets were able to form, allowing more giant planets to remain in stable orbits and having fewer interactions with the binaries that led to their ejections. This is also the case for less massive planets, e.g. of super-Earths and Neptune mass. With fewer massive planets forming, there are fewer opportunities for the planets to be forced on to eccentric orbits before being ejected from the system.

\begin{figure}
\centering
\includegraphics[scale=0.5]{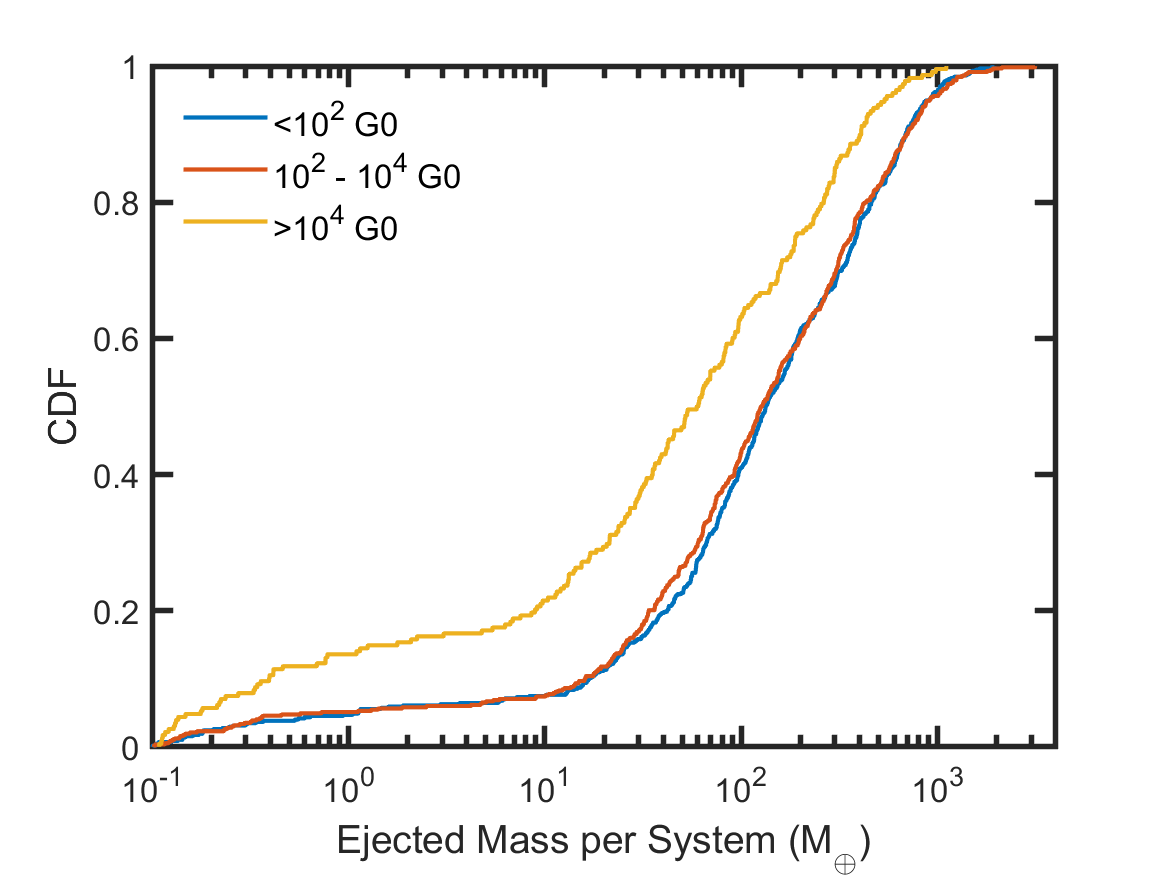}
\caption{Cumulative distribution functions for the total ejected planet mass from circumbinary systems evolving in different UV environments. The distributions show for weak UV environments ($<10^2 \rm {G_0}$, blue line), intermediate UV environments ($10^2$--$10^4 \rm {G_0}$, red line), and strong UV environments ($>10^4 \rm {G_0}$, yellow line).}
\label{fig:environment_mass}
\end{figure}

With Fig. \ref{fig:environment_triple} showing fewer planets being ejected from discs evolving in strong UV environments, we now look at the total mass of planets ejected from systems in different UV environments. Figure \ref{fig:environment_mass} shows the cumulative distributions of the total combined mass of planets ejected in weak ($<10^2 \rm G_0$, blue line), intermediate ($(10^2 {\rm G_0}$--$10^4 {\rm G_0})$, red line), and strong UV environments ($>10^4 \rm G_0$, yellow line). Further highlighting the similarities between the left-hand and middle panels in Fig. \ref{fig:environment_triple}, the red and blue lines in Fig. \ref{fig:environment_triple} are effectively identical. This shows that there is minimal effect on the planets forming and being ejected in weak and intermediate UV environments $(<10^{4}\rm G_0)$. Whilst this may show that weak and intermediate environments do not affect the formation of planets, it is worth noting that the protoplanetary discs they form in are affected by such weak environments, since the discs are truncated down to $\sim$few hundred $\au$ \citep{Haworth23,Coleman24MHD}. Whilst the weak and intermediate environments showed few differences in the total mass of planets ejected, those planets forming in stronger environments are on average less massive. Looking at the yellow line in Fig. \ref{fig:environment_mass}, 36\% of systems in strong UV environments ejected planets with a combined mass of $>100\me$. Comparing this value to the red and blue lines, this percentage increases to 57\% and 58\% respectively, further highlighting the increase in total planet mass ejected from those systems. Going to much less massive planets, 14\% of systems in strong UV environments ejected planets with a total mass below $1\me$, compared to $5\%$ in weak and intermediate environments.

As Figs. \ref{fig:environment_triple} and \ref{fig:environment_mass} show there are fewer planets ejected in strong UV environments, this should be an observable difference in observed star forming regions. With Fig. \ref{fig:EjectionTime} showing most ejections occur between 0.5--3 Myr, observations of young star forming regions with a variety of different UV field strengths, then there should be more planets, and larger total combined masses in weaker UV regions. Interestingly there is a negligible difference in the mass distribution of the planets ejected themselves, but there are just fewer planets that form and are ejected.

\begin{figure*}
\centering
\includegraphics[scale=0.5]{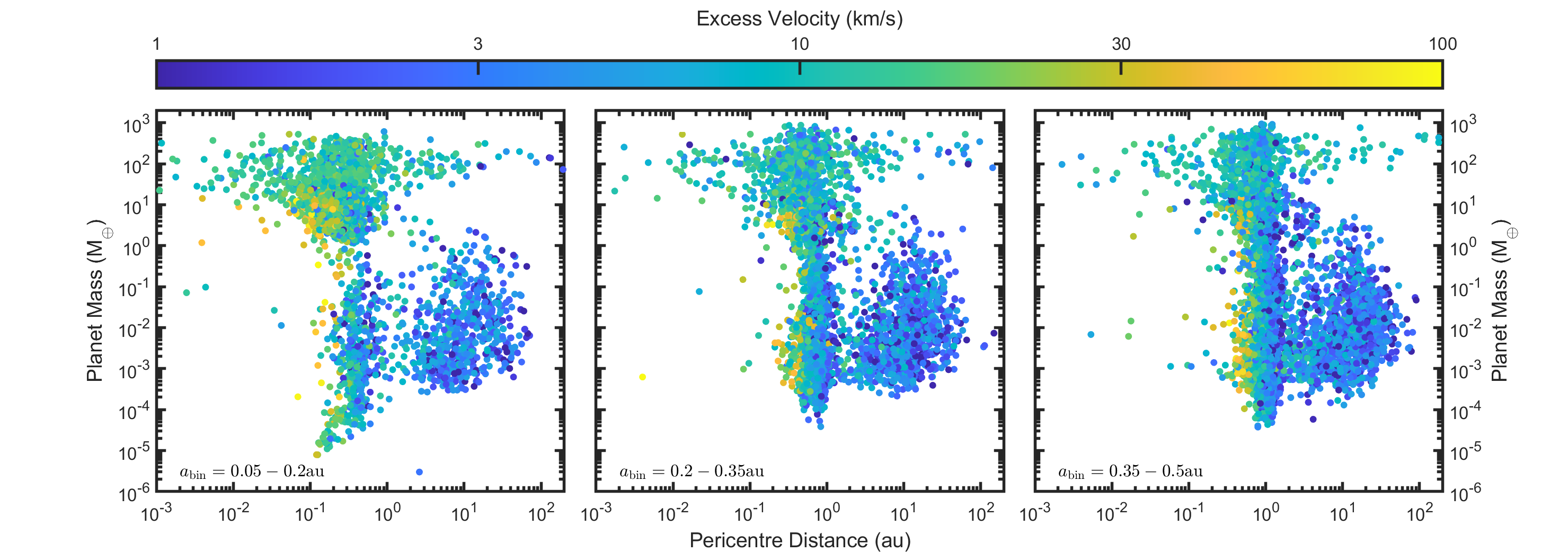}
\caption{Same as Fig. \ref{fig:alpha_triple}, but the panels show for planets ejected in systems with small binary separations ($0.05<a_{\rm bin}<0.2\au$, left-hand panel), intermediate binary separations ($0.2<a_{\rm bin}<0.35\au$, middle panel) and large binary separations ($0.35<a_{\rm bin}<0.5\au$, right-hand panel).}
\label{fig:separation_triple}
\end{figure*}

\subsubsection{Possibility of JUMBOs?}
Recent observations of the ONC cluster by JWST found numerous planetary mass objects, appearing unbound from nearby stellar systems \citep{Pearson23}. Interestingly, a significant fraction ($\sim 10\%$) of these free floating objects were found to be be of binary nature, and were termed as ``Jupiter-mass Binary Objects'' (JUMBOs). These objects with masses $0.5{\rm M_{Jup}}\le m_{\rm p} \le14 {\rm M_{Jup}}$, are typically found with separations $a_{\rm sep}<300 \au$. Whilst in our simulations, 18\% of systems ejected at least 2 planets with masses $m_{\rm p}>100\me$, it is extremely difficult to naturally form giant planets as binary objects, or even on coorbital orbits. Additionally it is even more unlikely to be able to eject both planets at a similar time, allowing them to remain gravitationally bound to each other.

By comparing the ejection time of planets in the same system, and additionally the distance between them, we can determine when planets are ejected either as binary objects, or fortuitously along similar trajectories. Unfortunately, nearly all of the giant planets ejected in our simulations leave the systems along vastly different trajectories. Indeed, the closest ``pair'' of giant planets ejected in the simulations, were separated by $\sim990\au$ when the second planet was ejected from the simulation and was 1000$\au$ from the central binary. Further showing the point of the large distances between planets ejected from the same system, looking at ``pairs'' of planets with masses $m_{\rm p}>10\me$, only 0.6\% of them had separations less than 1000$\au$. This shows how extremely rare it is for binary objects to be ejected from within circumbinary systems, pointing to other mechanisms for the formation of the observed JUMBO population. Such other mechanisms could include, formation in stellar systems and fortuitous ejection through flybys of nearby stars \citep{Wang24} or through star formation pathways \citep{PortegiesZwart24}.

\subsection{Effects of changing binary separations}

The final key input parameter that we test to explore the differences in the distributions of FFPs is the binary separation. Typically most works exploring the formation of circumbinary planets only use a specific set of binary parameters, including the binary separation \citep[e.g.][]{Coleman23,Coleman24}, since they are usually examining the formation of specific systems. Whilst we do not vary the binary mass ratio or eccentricity, since they affect the prescriptions for the central circumbinary cavity \citep{Thun17,Kley19}, we do vary the binary separation between 0.05--0.5 $\au$. Varying the binary separation has previously been unexplored in the formation of circumbinary planets, of which we will explore in future work. Here however, we examine the effects that the binary separation has on the properties and distributions of FFPs.

In Fig. \ref{fig:separation_triple} we again show three plots of planet mass versus the pericentre distance (ejection location) for planets forming in systems with different binary separations. The colours again show the excess velocities as the planets are ejected from the systems. The left-hand panel shows planets that formed in systems with separations $0.05\au\le a_{\rm bin}<0.2\au$, whilst the middle shows for $0.2\au\le a_{\rm bin}<0.35\au$, and the right-hand panel for $0.35\au\le a_{\rm bin}\le0.5\au$. Comparing the different panels, they all generally contain similar populations of planets, however there are some small and subtle differences. For the smallest binary separation (left-hand panel), there is a larger number of planets with masses between 1--100$\me$ ejected from around the cavity edges, situated between 0.1--1$\au$. Comparing this to the right-hand panel for the most separated binaries, the area of ejection is much more confined to a smaller region. Additionally, when comparing the excess velocity of these planets, it is clear that those ejected from the closer binaries in the left-hand panel, have larger velocities than their wider binary counterparts. This is not unsurprising since the orbital and escape velocities involved there are much larger, and so any deviations on to eccentric orbits, where the planets more freely interact with the stars, can lead to more energetic escapes. Furthermore, with the binaries being more compact, and ejections occurring in closer proximity, the likelihood of a planet on a hyperbolic orbit interacting with both binary stars, as well as other planets in the vicinity, also increases, which gives rise to a larger spread in the pericentre distance, since their velocities will be altered from the original ejecting event.

\begin{figure}
\centering
\includegraphics[scale=0.5]{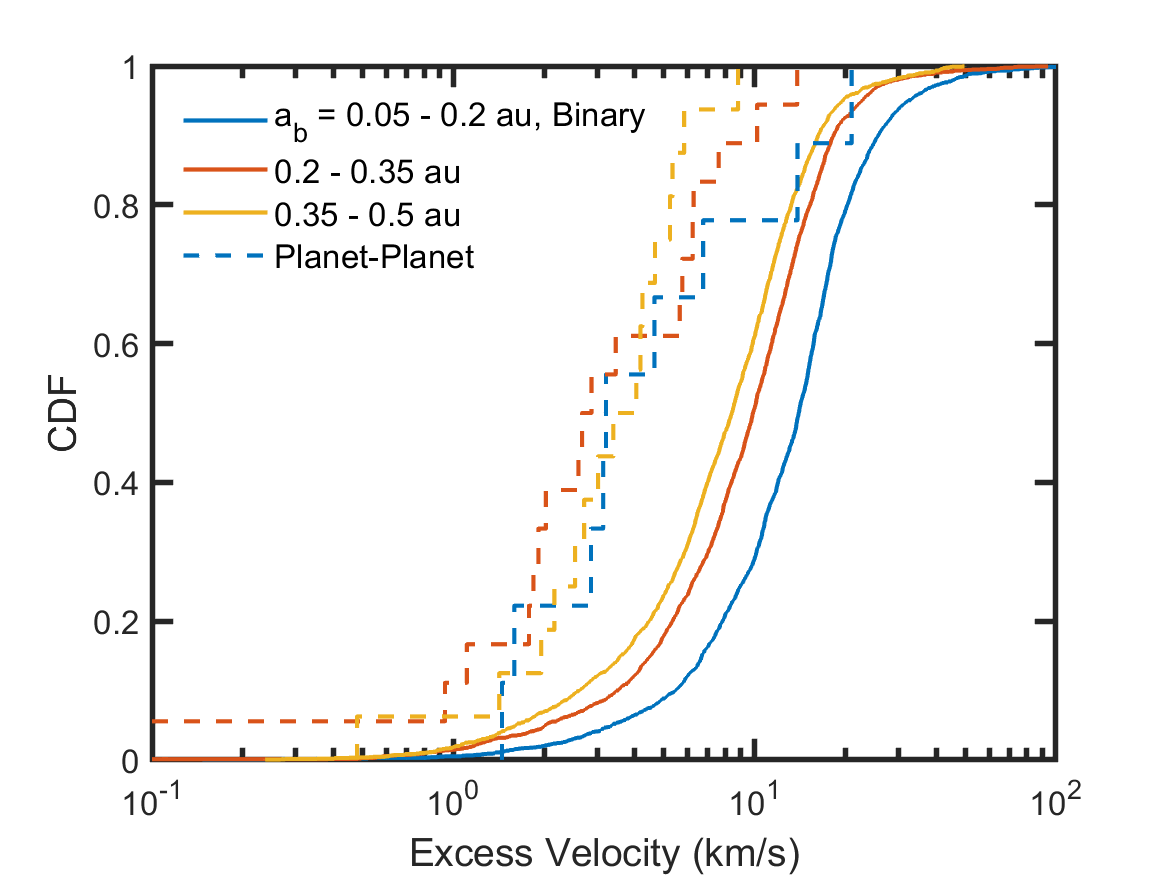}
\caption{Cumulative distribution functions of planet excess velocities for ejected planets with masses $m_{\rm p}>1\me$. The lines differentiate between planets ejected through interactions with the central binary (solid lines), and through planet--planet interactions (dashed lines). The colours show for planets ejected from systems with small binary separations ($0.05<a_{\rm bin}<0.2\au$, blue lines), intermediate binary separations ($0.2<a_{\rm bin}<0.35\au$, red lines) and large binary separations ($0.35<a_{\rm bin}<0.5\au$, yellow lines).}
\label{fig:separation_excess}
\end{figure}

Whilst the distributions of planet mass, number of planets ejected, and the total mass ejected, all remain similar as a function of the binary separation, there is one distribution that is significantly affected. As noticed in Fig. \ref{fig:separation_triple} the excess velocities that planets retained were larger for closer binaries than for wider binaries. Similar to Fig. \ref{fig:excess_velocity_stats}, we plot the cumulative distributions of excess velocities for ejected planets with masses $m_{\rm p}>1\me$ in Fig. \ref{fig:separation_excess}. Here we separate those planets ejected from the binary with solid lines to those through planet--planet interactions with dashed lines, whilst the colours show planets ejected from: close binaries (blue), intermediate binaries (red) and wider binaries (yellow). Whilst there appears to be little difference for those planets ejected through planet--planet interactions (though there are few planets ejected in this manner), there is a noticeable trend for those planets ejected through interactions with the binary stars. As the separation of the binary stars increases, the excess velocity distributions moves to lower values, approaching the planet--planet induced velocities. This would suggest that for even wider binaries, i.e. with separations of $\sim$few $\au$, there would be little difference in the excess velocity distributions irrespective of the ejection mechanism. This is unsurprising, since the escape velocity for the wider orbit planets would be reduced, and additionally, strong enough interactions with one of the binaries to eject the planet would be more easily attained, without the planet having extreme close encounters with one of the stars. With close binaries being observed frequently near to the sun \citep{Raghavan10} and with observing campaigns finding large numbers of eclipsing binaries \citep[e.g. {\it Kepler,}][]{Prsa11}, signatures in excess velocities should be readily observable in FFPs found in star-forming regions.

\section{Discussion and Conclusions}
\label{sec:conc}

In this work we have explored the formation of free floating planets (FFPs) within circumbinary systems, where the planets are initially able to form similarly to remaining bound planets, before they are ejected either through interactions with other planets or with the central binary stars themselves.
We used an updated version of the {\textit{N}}-body code \textsc{mercury6} including the effects of a central binary, and coupled to this a self-consistent 1D viscously evolving disc model containing prescriptions for planet migration, accretion of gaseous envelopes, pebble accretion, disc removal through photoevaporative winds and prescriptions taking into account the effects of the central binary such as an eccentric cavity.
Within the resulting populations we derived distributions and make observable predictions that can be tested with future observations.
We explored the effects of numerous initial parameters on the distributions of FFPs including, the strength of the local radiation environment, the level of turbulence in the disc, and binary separation.
The main results from our study can be summarised as follows.

(1) In agreement with previous works \citep[e.g.][]{Coleman23,Coleman24}, circumbinary systems are efficient factories of FFPs. On average, each binary system ejects between 2--7 planets with masses $m_{\rm p}>1\me$. When considering giant planets only, i.e. those with masses $m_{\rm p}>100\me$, then only 0.6 planets are ejected per system. Whilst the number of FFPs formed is consistent with other planet formation studies, they are reduced compared to preliminary observations in mircolensing surveys \citep{Sumi23}.

(2) Most of the FFPs that form within circumbinary system are ejected between 0.4--4 Myr, typically whist there are still circumbinary discs. This is important when attempting to observe FFPs in young star-forming regions, since they should be abundantly found there.

(3) Comparing the mass distributions of ejected planets to those remaining bound and orbiting close to the central stars, we find that for planets with masses $m_{\rm p}>20\me$, the distributions are similar. This means that observed mass distributions and occurrence rates from microlensing surveys should be comparable to the true populations of planets in circumbinary systems, close to the stars, and additionally to other observation surveys, e.g. transit or radial velocity, of circumbinary systems which are generally biased towards planets orbiting close to the central stars.

(4) Along with the mass distributions of FFPs, another main observable has arisen within the population of FFPs. As the planets are ejected from the systems, they retain significant excess velocities, between 8--16 $\rm kms^{-1}$. This is much larger than observed velocity dispersions of stars in local star forming regions \citep{VanAltena88,Kim19,Theissen23}, and so determinations of the velocity dispersions of FFPs in stellar clusters should differentiate whether those planets arose from either star or planet formation process. 

(5) Additionally the velocity dispersions of FFPs ejected through interactions with the binary stars is $\sim$3 times larger than those ejected through planet--planet interactions. This means that observed velocity dispersions can determine whether binary or single star systems are the dominant origin of FFPs. From our simulations, we find that planet--planet interactions only account for 0.6\% of all FFPs larger than 1 $\me$, indicating that binary systems are the most likely origin of FFPs when taking observational constraints on binary fractions into account \citep{Offner23}.

(6) The initial parameters studied also affect the distributions of the properties of FFPs. Notably, the level of turbulence affects the number of planets ejected, and their accompanying mass distributions. Weaker level of turbulence favours larger numbers of planets ejected, with more less massive planets (96\% are less massive than 100$\me$), with stronger turbulence favouring more massive planets (46\% with masses greater than 100$\me$). The strength of the local environment generally affects the total mass of planets ejected from circumbinary systems, with those in strong UV environments being most severely affected. Both of these effects of turbulence and the environment, are due to the reduction in solid material available for accretion by planets, either through faster dispersal of the disc, or through extended pebble and dust layers that hinder planetesimal formation and pebble accretion.

(7) Finally, the separation of the binary is found to have little effect on the frequency of FFPs, and their resultant mass distributions. This is due to the main planet formation processes occurring far from the influence of the binary stars, and then migration transporting the planets to the vicinity of the binaries. The binary separation does however, affect the velocity distributions of ejected planets, with wider binaries causing a reduction in the average excess velocity since planets are more easily ejected without having to be deflected onto orbits that pass close to the individual central stars.\\

The simulations here show that it is possible to use observations of the distributions of FFPs to determine their origins, as well as further properties of the environment, both the discs and the star forming regions they formed in. Differences in the distributions of FFP masses, their frequencies, and excess velocities, can all indicate whether single stars or circumbinary systems are the fundamental birthplace of FFPs.
However, whilst this work contains numerous simulations, and explores a broad parameter space, it does not constitute a full population of forming circumbinary systems. Such a population would include varying combined stellar masses, mass ratios, and binary eccentricities, that would be motivated by observations. Since the modelling of central cavities in circumbinary discs require specific prescriptions, depending on the binary properties, it is not currently feasible to derive such a population. Should such a population be performed in future work, then comparisons between that population and observed populations would give even more valuable insight into the formation of these intriguing objects.

In addition, planet formation models are constantly evolving and adding new physics. With the population of FFPs originating in different regions of the circumbinary discs, then they would accrete material with varied compositions, both in the gas, and in the solids. In future work we will incorporate compositional models \citep[e.g.][]{Thiabaud14,Thiabaud15} that will allow us to determine if there are any chemical signatures within the FFP populations that can point to not only the stellar environment in which they formed, but more tightly constrain the regions within their natal protoplanetary discs. Should spectroscopic observations be performed for known directly imaged FFPs, and should we know their origins, then this would greatly inform on planet formation models as a whole, and indicate where improvements are needed within our models. Only then, will we be able to develop a full and complete model of planet formation that can explain the populations observed today, both the FFP population, and those exoplanets found through other methods, e.g. transit or radial velocity surveys.

\section*{Data Availability}
The data underlying this article will be shared on reasonable request to the corresponding author.

\section*{Acknowledgements}
GALC acknowledges funding from the Royal Society under the Dorothy Hodgkin Fellowship of T. J. Haworth, STFC through grant ST/P000592/1, and the Leverhulme Trust through grant RPG-2018-418.
This research utilised Queen Mary's Apocrita HPC facility, supported by QMUL Research-IT (http://doi.org/10.5281/zenodo.438045).
This work was performed using the Cambridge Service for Data Driven Discovery (CSD3), part of which is operated by the University of Cambridge Research Computing on behalf of the STFC DiRAC HPC Facility (www.dirac.ac.uk). The DiRAC component of CSD3 was funded by BEIS capital funding via STFC capital grants ST/P002307/1 and ST/R002452/1 and STFC operations grant ST/R00689X/1. DiRAC is part of the National e-Infrastructure.
The authors would like to acknowledge the support provided by the GridPP Collaboration, in particular from the Queen Mary University of London Tier two centre.

\bibliographystyle{mnras}
\bibliography{references}{}

\begin{thebibliography}{}
\makeatletter
\relax
\def\mn@urlcharsother{\let\do\@makeother \do\$\do\&\do\#\do\^\do\_\do\%\do\~}
\def\mn@doi{\begingroup\mn@urlcharsother \@ifnextchar [ {\mn@doi@} {\mn@doi@[]}}
\def\mn@doi@[#1]#2{\def\@tempa{#1}\ifx\@tempa\@empty \href {http://dx.doi.org/#2} {doi:#2}\else \href {http://dx.doi.org/#2} {#1}\fi \endgroup}
\def\mn@eprint#1#2{\mn@eprint@#1:#2::\@nil}
\def\mn@eprint@arXiv#1{\href {http://arxiv.org/abs/#1} {{\tt arXiv:#1}}}
\def\mn@eprint@dblp#1{\href {http://dblp.uni-trier.de/rec/bibtex/#1.xml} {dblp:#1}}
\def\mn@eprint@#1:#2:#3:#4\@nil{\def\@tempa {#1}\def\@tempb {#2}\def\@tempc {#3}\ifx \@tempc \@empty \let \@tempc \@tempb \let \@tempb \@tempa \fi \ifx \@tempb \@empty \def\@tempb {arXiv}\fi \@ifundefined {mn@eprint@\@tempb}{\@tempb:\@tempc}{\expandafter \expandafter \csname mn@eprint@\@tempb\endcsname \expandafter{\@tempc}}}

\bibitem[\protect\citeauthoryear{{Abod}, {Simon}, {Li}, {Armitage}, {Youdin}  \& {Kretke}}{{Abod} et~al.}{2019}]{Abod19}
{Abod} C.~P.,  {Simon} J.~B.,  {Li} R.,  {Armitage} P.~J.,  {Youdin} A.~N.,   {Kretke} K.~A.,  2019, \mn@doi [\apj] {10.3847/1538-4357/ab40a3}, \href {https://ui.adsabs.harvard.edu/abs/2019ApJ...883..192A} {883, 192}

\bibitem[\protect\citeauthoryear{{Andrews}, {Wilner}, {Hughes}, {Qi}  \& {Dullemond}}{{Andrews} et~al.}{2010}]{Andrews10}
{Andrews} S.~M.,  {Wilner} D.~J.,  {Hughes} A.~M.,  {Qi} C.,   {Dullemond} C.~P.,  2010, \mn@doi [\apj] {10.1088/0004-637X/723/2/1241}, \href {https://ui.adsabs.harvard.edu/abs/2010ApJ...723.1241A} {723, 1241}

\bibitem[\protect\citeauthoryear{{Artymowicz} \& {Lubow}}{{Artymowicz} \& {Lubow}}{1994}]{Artymowicz94}
{Artymowicz} P.,  {Lubow} S.~H.,  1994, \mn@doi [\apj] {10.1086/173679}, \href {https://ui.adsabs.harvard.edu/abs/1994ApJ...421..651A} {421, 651}

\bibitem[\protect\citeauthoryear{{Ataiee}, {Baruteau}, {Alibert}  \& {Benz}}{{Ataiee} et~al.}{2018}]{Ataiee18}
{Ataiee} S.,  {Baruteau} C.,  {Alibert} Y.,   {Benz} W.,  2018, \mn@doi [\aap] {10.1051/0004-6361/201732026}, \href {https://ui.adsabs.harvard.edu/abs/2018A&A...615A.110A} {615, A110}

\bibitem[\protect\citeauthoryear{{Ben{\'\i}tez-Llambay} \& {Masset}}{{Ben{\'\i}tez-Llambay} \& {Masset}}{2016}]{FARGO-3D-2016}
{Ben{\'\i}tez-Llambay} P.,  {Masset} F.~S.,  2016, \mn@doi [\apjs] {10.3847/0067-0049/223/1/11}, \href {https://ui.adsabs.harvard.edu/abs/2016ApJS..223...11B} {223, 11}

\bibitem[\protect\citeauthoryear{{Bennett} et~al.,}{{Bennett} et~al.}{2018}]{Bennett18}
{Bennett} D.~P.,  et~al., 2018, arXiv e-prints, \href {https://ui.adsabs.harvard.edu/abs/2018arXiv180308564B} {p. arXiv:1803.08564}

\bibitem[\protect\citeauthoryear{{Bitsch}, {Morbidelli}, {Johansen}, {Lega}, {Lambrechts}  \& {Crida}}{{Bitsch} et~al.}{2018}]{Bitsch18}
{Bitsch} B.,  {Morbidelli} A.,  {Johansen} A.,  {Lega} E.,  {Lambrechts} M.,   {Crida} A.,  2018, \mn@doi [\aap] {10.1051/0004-6361/201731931}, \href {https://ui.adsabs.harvard.edu/abs/2018A&A...612A..30B} {612, A30}

\bibitem[\protect\citeauthoryear{{Chambers}}{{Chambers}}{1999}]{Chambers}
{Chambers} J.~E.,  1999, \mn@doi [\mnras] {10.1046/j.1365-8711.1999.02379.x}, \href {http://adsabs.harvard.edu/abs/1999MNRAS.304..793C} {304, 793}

\bibitem[\protect\citeauthoryear{{Chambers}, {Quintana}, {Duncan}  \& {Lissauer}}{{Chambers} et~al.}{2002}]{ChambersBinary}
{Chambers} J.~E.,  {Quintana} E.~V.,  {Duncan} M.~J.,   {Lissauer} J.~J.,  2002, \mn@doi [\aj] {10.1086/340074}, \href {https://ui.adsabs.harvard.edu/abs/2002AJ....123.2884C} {123, 2884}

\bibitem[\protect\citeauthoryear{{Chen}, {Martin}, {Lubow}  \& {Nixon}}{{Chen} et~al.}{2024}]{Cheng24}
{Chen} C.,  {Martin} R.~G.,  {Lubow} S.~H.,   {Nixon} C.~J.,  2024, \mn@doi [\apjl] {10.3847/2041-8213/ad17c5}, \href {https://ui.adsabs.harvard.edu/abs/2024ApJ...961L...5C} {961, L5}

\bibitem[\protect\citeauthoryear{{Clarke}, {Gendrin}  \& {Sotomayor}}{{Clarke} et~al.}{2001}]{Clarke2001}
{Clarke} C.~J.,  {Gendrin} A.,   {Sotomayor} M.,  2001, \mn@doi [\mnras] {10.1046/j.1365-8711.2001.04891.x}, \href {http://adsabs.harvard.edu/abs/2001MNRAS.328..485C} {328, 485}

\bibitem[\protect\citeauthoryear{{Coleman}}{{Coleman}}{2021}]{Coleman21}
{Coleman} G. A.~L.,  2021, \mn@doi [\mnras] {10.1093/mnras/stab1904}, \href {https://ui.adsabs.harvard.edu/abs/2021MNRAS.506.3596C} {506, 3596}

\bibitem[\protect\citeauthoryear{{Coleman} \& {Haworth}}{{Coleman} \& {Haworth}}{2022}]{Coleman22}
{Coleman} G. A.~L.,  {Haworth} T.~J.,  2022, \mn@doi [MNRAS] {10.1093/mnras/stac1513}, \href {https://ui.adsabs.harvard.edu/abs/2022MNRAS.514.2315C} {514, 2315}

\bibitem[\protect\citeauthoryear{{Coleman} \& {Nelson}}{{Coleman} \& {Nelson}}{2014}]{ColemanNelson14}
{Coleman} G.~A.~L.,  {Nelson} R.~P.,  2014, \mn@doi [\mnras] {10.1093/mnras/stu1715}, \href {http://adsabs.harvard.edu/abs/2014MNRAS.445..479C} {445, 479}

\bibitem[\protect\citeauthoryear{{Coleman} \& {Nelson}}{{Coleman} \& {Nelson}}{2016a}]{ColemanNelson16}
{Coleman} G.~A.~L.,  {Nelson} R.~P.,  2016a, \mn@doi [\mnras] {10.1093/mnras/stw149}, \href {http://adsabs.harvard.edu/abs/2016MNRAS.457.2480C} {457, 2480}

\bibitem[\protect\citeauthoryear{{Coleman} \& {Nelson}}{{Coleman} \& {Nelson}}{2016b}]{ColemanNelson16b}
{Coleman} G.~A.~L.,  {Nelson} R.~P.,  2016b, \mn@doi [\mnras] {10.1093/mnras/stw1177}, \href {http://adsabs.harvard.edu/abs/2016MNRAS.460.2779C} {460, 2779}

\bibitem[\protect\citeauthoryear{{Coleman}, {Papaloizou}  \& {Nelson}}{{Coleman} et~al.}{2017}]{CPN17}
{Coleman} G.~A.~L.,  {Papaloizou} J.~C.~B.,   {Nelson} R.~P.,  2017, \mn@doi [\mnras] {10.1093/mnras/stx1297}, \href {http://adsabs.harvard.edu/abs/2017MNRAS.470.3206C} {470, 3206}

\bibitem[\protect\citeauthoryear{{Coleman}, {Nelson}  \& {Triaud}}{{Coleman} et~al.}{2022}]{Coleman22b}
{Coleman} G. A.~L.,  {Nelson} R.~P.,   {Triaud} A. H.~M.~J.,  2022, \mn@doi [\mnras] {10.1093/mnras/stac1029}, \href {https://ui.adsabs.harvard.edu/abs/2022MNRAS.513.2563C} {513, 2563}

\bibitem[\protect\citeauthoryear{{Coleman}, {Nelson}  \& {Triaud}}{{Coleman} et~al.}{2023}]{Coleman23}
{Coleman} G. A.~L.,  {Nelson} R.~P.,   {Triaud} A. H.~M.~J.,  2023, \mn@doi [\mnras] {10.1093/mnras/stad833}, \href {https://ui.adsabs.harvard.edu/abs/2023MNRAS.522.4352C} {522, 4352}

\bibitem[\protect\citeauthoryear{{Coleman}, {Nelson}, {Triaud}  \& {Standing}}{{Coleman} et~al.}{2024a}]{Coleman24}
{Coleman} G. A.~L.,  {Nelson} R.~P.,  {Triaud} A. H.~M.~J.,   {Standing} M.~R.,  2024a, \mn@doi [\mnras] {10.1093/mnras/stad3216}, \href {https://ui.adsabs.harvard.edu/abs/2024MNRAS.527..414C} {527, 414}

\bibitem[\protect\citeauthoryear{{Coleman}, {Mroueh}  \& {Haworth}}{{Coleman} et~al.}{2024b}]{Coleman24MHD}
{Coleman} G. A.~L.,  {Mroueh} J.~K.,   {Haworth} T.~J.,  2024b, \mn@doi [\mnras] {10.1093/mnras/stad3692}, \href {https://ui.adsabs.harvard.edu/abs/2024MNRAS.527.7588C} {527, 7588}

\bibitem[\protect\citeauthoryear{{Cresswell} \& {Nelson}}{{Cresswell} \& {Nelson}}{2008}]{cressnels}
{Cresswell} P.,  {Nelson} R.~P.,  2008, \mn@doi [\aap] {10.1051/0004-6361:20079178}, \href {http://adsabs.harvard.edu/abs/2008A%26A...482..677C} {482, 677}

\bibitem[\protect\citeauthoryear{{Crida}, {Morbidelli}  \& {Masset}}{{Crida} et~al.}{2006}]{Crida}
{Crida} A.,  {Morbidelli} A.,   {Masset} F.,  2006, \mn@doi [\icarus] {10.1016/j.icarus.2005.10.007}, \href {http://adsabs.harvard.edu/abs/2006Icar..181..587C} {181, 587}

\bibitem[\protect\citeauthoryear{{Doyle} et~al.,}{{Doyle} et~al.}{2011}]{Doyle11}
{Doyle} L.~R.,  et~al., 2011, \mn@doi [Science] {10.1126/science.1210923}, \href {https://ui.adsabs.harvard.edu/abs/2011Sci...333.1602D} {333, 1602}

\bibitem[\protect\citeauthoryear{{Dutrey}, {Guilloteau}  \& {Simon}}{{Dutrey} et~al.}{1994}]{Dutrey94}
{Dutrey} A.,  {Guilloteau} S.,   {Simon} M.,  1994, \aap, \href {https://ui.adsabs.harvard.edu/abs/1994A&A...286..149D} {286, 149}

\bibitem[\protect\citeauthoryear{{Ercolano}, {Picogna}, {Monsch}, {Drake}  \& {Preibisch}}{{Ercolano} et~al.}{2021}]{Ercolano21}
{Ercolano} B.,  {Picogna} G.,  {Monsch} K.,  {Drake} J.~J.,   {Preibisch} T.,  2021, \mn@doi [\mnras] {10.1093/mnras/stab2590}, \href {https://ui.adsabs.harvard.edu/abs/2021MNRAS.508.1675E} {508, 1675}

\bibitem[\protect\citeauthoryear{{Fendyke} \& {Nelson}}{{Fendyke} \& {Nelson}}{2014}]{Fendyke}
{Fendyke} S.~M.,  {Nelson} R.~P.,  2014, \mn@doi [\mnras] {10.1093/mnras/stt1867}, \href {http://adsabs.harvard.edu/abs/2014MNRAS.437...96F} {437, 96}

\bibitem[\protect\citeauthoryear{{Flaherty} et~al.,}{{Flaherty} et~al.}{2017}]{Flaherty17}
{Flaherty} K.~M.,  et~al., 2017, \mn@doi [\apj] {10.3847/1538-4357/aa79f9}, \href {https://ui.adsabs.harvard.edu/abs/2017ApJ...843..150F} {843, 150}

\bibitem[\protect\citeauthoryear{{Flaischlen}, {Preibisch}, {Manara}  \& {Ercolano}}{{Flaischlen} et~al.}{2021}]{Flaischlen21}
{Flaischlen} S.,  {Preibisch} T.,  {Manara} C.~F.,   {Ercolano} B.,  2021, \mn@doi [\aap] {10.1051/0004-6361/202039746}, \href {https://ui.adsabs.harvard.edu/abs/2021A&A...648A.121F} {648, A121}

\bibitem[\protect\citeauthoryear{{Fortier}, {Alibert}, {Carron}, {Benz}  \& {Dittkrist}}{{Fortier} et~al.}{2013}]{Fortier13}
{Fortier} A.,  {Alibert} Y.,  {Carron} F.,  {Benz} W.,   {Dittkrist} K.~M.,  2013, \mn@doi [\aap] {10.1051/0004-6361/201220241}, \href {https://ui.adsabs.harvard.edu/abs/2013A&A...549A..44F} {549, A44}

\bibitem[\protect\citeauthoryear{{Habing}}{{Habing}}{1968}]{Habing68}
{Habing} H.~J.,  1968, \bain, \href {https://ui.adsabs.harvard.edu/abs/1968BAN....19..421H} {19, 421}

\bibitem[\protect\citeauthoryear{{Haworth}, {Clarke}, {Rahman}, {Winter}  \& {Facchini}}{{Haworth} et~al.}{2018}]{Haworth18}
{Haworth} T.~J.,  {Clarke} C.~J.,  {Rahman} W.,  {Winter} A.~J.,   {Facchini} S.,  2018, \mn@doi [\mnras] {10.1093/mnras/sty2323}, \href {https://ui.adsabs.harvard.edu/abs/2018MNRAS.481..452H} {481, 452}

\bibitem[\protect\citeauthoryear{{Haworth}, {Cadman}, {Meru}, {Hall}, {Albertini}, {Forgan}, {Rice}  \& {Owen}}{{Haworth} et~al.}{2020}]{Haworth20}
{Haworth} T.~J.,  {Cadman} J.,  {Meru} F.,  {Hall} C.,  {Albertini} E.,  {Forgan} D.,  {Rice} K.,   {Owen} J.~E.,  2020, \mn@doi [\mnras] {10.1093/mnras/staa883}, \href {https://ui.adsabs.harvard.edu/abs/2020MNRAS.494.4130H} {494, 4130}

\bibitem[\protect\citeauthoryear{{Haworth}, {Coleman}, {Qiao}, {Sellek}  \& {Askari}}{{Haworth} et~al.}{2023}]{Haworth23}
{Haworth} T.~J.,  {Coleman} G. A.~L.,  {Qiao} L.,  {Sellek} A.~D.,   {Askari} K.,  2023, \mn@doi [\mnras] {10.1093/mnras/stad3054}, \href {https://ui.adsabs.harvard.edu/abs/2023MNRAS.526.4315H} {526, 4315}

\bibitem[\protect\citeauthoryear{{Isella}, {Carpenter}  \& {Sargent}}{{Isella} et~al.}{2009}]{Isella09}
{Isella} A.,  {Carpenter} J.~M.,   {Sargent} A.~I.,  2009, \mn@doi [\apj] {10.1088/0004-637X/701/1/260}, \href {https://ui.adsabs.harvard.edu/abs/2009ApJ...701..260I} {701, 260}

\bibitem[\protect\citeauthoryear{{Ivezi{\'c}} et~al.,}{{Ivezi{\'c}} et~al.}{2019}]{LSST_2019}
{Ivezi{\'c}} {\v{Z}}.,  et~al., 2019, \mn@doi [\apj] {10.3847/1538-4357/ab042c}, \href {https://ui.adsabs.harvard.edu/abs/2019ApJ...873..111I} {873, 111}

\bibitem[\protect\citeauthoryear{{Johansen} \& {Lambrechts}}{{Johansen} \& {Lambrechts}}{2017}]{Johansen17}
{Johansen} A.,  {Lambrechts} M.,  2017, \mn@doi [Annual Review of Earth and Planetary Sciences] {10.1146/annurev-earth-063016-020226}, \href {http://adsabs.harvard.edu/abs/2017AREPS..45..359J} {45, 359}

\bibitem[\protect\citeauthoryear{{Johansen}, {Oishi}, {Mac Low}, {Klahr}, {Henning}  \& {Youdin}}{{Johansen} et~al.}{2007}]{Johansen07}
{Johansen} A.,  {Oishi} J.~S.,  {Mac Low} M.-M.,  {Klahr} H.,  {Henning} T.,   {Youdin} A.,  2007, \mn@doi [\nat] {10.1038/nature06086}, \href {http://adsabs.harvard.edu/abs/2007Natur.448.1022J} {448, 1022}

\bibitem[\protect\citeauthoryear{{Johansen}, {Youdin}  \& {Klahr}}{{Johansen} et~al.}{2009}]{JohansenYoudin2009}
{Johansen} A.,  {Youdin} A.,   {Klahr} H.,  2009, \mn@doi [\apj] {10.1088/0004-637X/697/2/1269}, \href {http://adsabs.harvard.edu/abs/2009ApJ...697.1269J} {697, 1269}

\bibitem[\protect\citeauthoryear{{Johansen}, {Mac Low}, {Lacerda}  \& {Bizzarro}}{{Johansen} et~al.}{2015}]{Johansen15}
{Johansen} A.,  {Mac Low} M.-M.,  {Lacerda} P.,   {Bizzarro} M.,  2015, \mn@doi [Science Advances] {10.1126/sciadv.1500109}, \href {https://ui.adsabs.harvard.edu/abs/2015SciA....1E0109J} {1, 1500109}

\bibitem[\protect\citeauthoryear{{Kim}, {Lu}, {Konopacky}, {Chu}, {Toller}, {Anderson}, {Theissen}  \& {Morris}}{{Kim} et~al.}{2019}]{Kim19}
{Kim} D.,  {Lu} J.~R.,  {Konopacky} Q.,  {Chu} L.,  {Toller} E.,  {Anderson} J.,  {Theissen} C.~A.,   {Morris} M.~R.,  2019, \mn@doi [\aj] {10.3847/1538-3881/aafb09}, \href {https://ui.adsabs.harvard.edu/abs/2019AJ....157..109K} {157, 109}

\bibitem[\protect\citeauthoryear{{Kley}, {Thun}  \& {Penzlin}}{{Kley} et~al.}{2019}]{Kley19}
{Kley} W.,  {Thun} D.,   {Penzlin} A. B.~T.,  2019, \mn@doi [\aap] {10.1051/0004-6361/201935503}, \href {https://ui.adsabs.harvard.edu/abs/2019A&A...627A..91K} {627, A91}

\bibitem[\protect\citeauthoryear{{Koshimoto} et~al.,}{{Koshimoto} et~al.}{2023}]{Koshimoto23}
{Koshimoto} N.,  et~al., 2023, \mn@doi [\aj] {10.3847/1538-3881/ace689}, \href {https://ui.adsabs.harvard.edu/abs/2023AJ....166..107K} {166, 107}

\bibitem[\protect\citeauthoryear{{Kostov} et~al.,}{{Kostov} et~al.}{2020}]{Kostov20}
{Kostov} V.~B.,  et~al., 2020, \mn@doi [\aj] {10.3847/1538-3881/ab8a48}, \href {https://ui.adsabs.harvard.edu/abs/2020AJ....159..253K} {159, 253}

\bibitem[\protect\citeauthoryear{{Kroupa}}{{Kroupa}}{2001}]{Kroupa01}
{Kroupa} P.,  2001, \mn@doi [\mnras] {10.1046/j.1365-8711.2001.04022.x}, \href {https://ui.adsabs.harvard.edu/abs/2001MNRAS.322..231K} {322, 231}

\bibitem[\protect\citeauthoryear{{Lambrechts} \& {Johansen}}{{Lambrechts} \& {Johansen}}{2012}]{Lambrechts12}
{Lambrechts} M.,  {Johansen} A.,  2012, \mn@doi [\aap] {10.1051/0004-6361/201219127}, \href {http://adsabs.harvard.edu/abs/2012A%26A...544A..32L} {544, A32}

\bibitem[\protect\citeauthoryear{{Lambrechts} \& {Johansen}}{{Lambrechts} \& {Johansen}}{2014}]{Lambrechts14}
{Lambrechts} M.,  {Johansen} A.,  2014, \mn@doi [\aap] {10.1051/0004-6361/201424343}, \href {http://adsabs.harvard.edu/abs/2014A%26A...572A.107L} {572, A107}

\bibitem[\protect\citeauthoryear{{Lin} \& {Papaloizou}}{{Lin} \& {Papaloizou}}{1986}]{LinPapaloizou86}
{Lin} D.~N.~C.,  {Papaloizou} J.,  1986, \mn@doi [\apj] {10.1086/164653}, \href {http://adsabs.harvard.edu/abs/1986ApJ...309..846L} {309, 846}

\bibitem[\protect\citeauthoryear{{Lynden-Bell} \& {Pringle}}{{Lynden-Bell} \& {Pringle}}{1974}]{Lynden-BellPringle1974}
{Lynden-Bell} D.,  {Pringle} J.~E.,  1974, \mnras, \href {http://adsabs.harvard.edu/abs/1974MNRAS.168..603L} {168, 603}

\bibitem[\protect\citeauthoryear{{Ma}, {Mao}, {Ida}, {Zhu}  \& {Lin}}{{Ma} et~al.}{2016}]{Ma16}
{Ma} S.,  {Mao} S.,  {Ida} S.,  {Zhu} W.,   {Lin} D. N.~C.,  2016, \mn@doi [\mnras] {10.1093/mnrasl/slw110}, \href {https://ui.adsabs.harvard.edu/abs/2016MNRAS.461L.107M} {461, L107}

\bibitem[\protect\citeauthoryear{{Miret-Roig} et~al.,}{{Miret-Roig} et~al.}{2022}]{MiretRoig22}
{Miret-Roig} N.,  et~al., 2022, \mn@doi [Nature Astronomy] {10.1038/s41550-021-01513-x}, \href {https://ui.adsabs.harvard.edu/abs/2022NatAs...6...89M} {6, 89}

\bibitem[\protect\citeauthoryear{{Mr{\'o}z} et~al.,}{{Mr{\'o}z} et~al.}{2019}]{Mroz19}
{Mr{\'o}z} P.,  et~al., 2019, \mn@doi [\aap] {10.1051/0004-6361/201834557}, \href {https://ui.adsabs.harvard.edu/abs/2019A&A...622A.201M} {622, A201}

\bibitem[\protect\citeauthoryear{{Mr{\'o}z} et~al.,}{{Mr{\'o}z} et~al.}{2020}]{Mroz20}
{Mr{\'o}z} P.,  et~al., 2020, \mn@doi [\apjl] {10.3847/2041-8213/abbfad}, \href {https://ui.adsabs.harvard.edu/abs/2020ApJ...903L..11M} {903, L11}

\bibitem[\protect\citeauthoryear{{Mutter}, {Pierens}  \& {Nelson}}{{Mutter} et~al.}{2017}]{Mutter17D}
{Mutter} M.~M.,  {Pierens} A.,   {Nelson} R.~P.,  2017, \mn@doi [\mnras] {10.1093/mnras/stw2768}, \href {https://ui.adsabs.harvard.edu/abs/2017MNRAS.465.4735M} {465, 4735}

\bibitem[\protect\citeauthoryear{{Nelson}}{{Nelson}}{2003}]{Nelson03}
{Nelson} R.~P.,  2003, \mn@doi [\mnras] {10.1046/j.1365-8711.2003.06929.x}, \href {https://ui.adsabs.harvard.edu/abs/2003MNRAS.345..233N} {345, 233}

\bibitem[\protect\citeauthoryear{{Offner}, {Moe}, {Kratter}, {Sadavoy}, {Jensen}  \& {Tobin}}{{Offner} et~al.}{2023}]{Offner23}
{Offner} S.~S.~R.,  {Moe} M.,  {Kratter} K.~M.,  {Sadavoy} S.~I.,  {Jensen} E.~L.~N.,   {Tobin} J.~J.,  2023, in {Inutsuka} S.,  {Aikawa} Y.,  {Muto} T.,  {Tomida} K.,   {Tamura} M.,  eds,  Astronomical Society of the Pacific Conference Series Vol. 534, Protostars and Planets VII. p.~275 (\mn@eprint {arXiv} {2203.10066}), \mn@doi{10.48550/arXiv.2203.10066}

\bibitem[\protect\citeauthoryear{{Owen}, {Ercolano}, {Clarke}  \& {Alexander}}{{Owen} et~al.}{2010}]{Owen10}
{Owen} J.~E.,  {Ercolano} B.,  {Clarke} C.~J.,   {Alexander} R.~D.,  2010, \mn@doi [\mnras] {10.1111/j.1365-2966.2009.15771.x}, \href {https://ui.adsabs.harvard.edu/abs/2010MNRAS.401.1415O} {401, 1415}

\bibitem[\protect\citeauthoryear{{Paardekooper}, {Baruteau}, {Crida}  \& {Kley}}{{Paardekooper} et~al.}{2010}]{pdk10}
{Paardekooper} S.-J.,  {Baruteau} C.,  {Crida} A.,   {Kley} W.,  2010, \mn@doi [\mnras] {10.1111/j.1365-2966.2009.15782.x}, \href {http://adsabs.harvard.edu/abs/2010MNRAS.401.1950P} {401, 1950}

\bibitem[\protect\citeauthoryear{{Paardekooper}, {Baruteau}  \& {Kley}}{{Paardekooper} et~al.}{2011}]{pdk11}
{Paardekooper} S.-J.,  {Baruteau} C.,   {Kley} W.,  2011, \mn@doi [\mnras] {10.1111/j.1365-2966.2010.17442.x}, \href {http://adsabs.harvard.edu/abs/2011MNRAS.410..293P} {410, 293}

\bibitem[\protect\citeauthoryear{{Padoan} \& {Nordlund}}{{Padoan} \& {Nordlund}}{2002}]{Padoan02}
{Padoan} P.,  {Nordlund} {\r{A}}.,  2002, \mn@doi [\apj] {10.1086/341790}, \href {https://ui.adsabs.harvard.edu/abs/2002ApJ...576..870P} {576, 870}

\bibitem[\protect\citeauthoryear{{Papaloizou} \& {Nelson}}{{Papaloizou} \& {Nelson}}{2005}]{PapNelson2005}
{Papaloizou} J.~C.~B.,  {Nelson} R.~P.,  2005, \mn@doi [\aap] {10.1051/0004-6361:20042029}, \href {http://adsabs.harvard.edu/abs/2005A%26A...433..247P} {433, 247}

\bibitem[\protect\citeauthoryear{{Papaloizou} \& {Terquem}}{{Papaloizou} \& {Terquem}}{1999}]{Pap-Terquem-envelopes}
{Papaloizou} J.~C.~B.,  {Terquem} C.,  1999, \mn@doi [\apj] {10.1086/307581}, \href {http://adsabs.harvard.edu/abs/1999ApJ...521..823P} {521, 823}

\bibitem[\protect\citeauthoryear{{Pearson} \& {McCaughrean}}{{Pearson} \& {McCaughrean}}{2023}]{Pearson23}
{Pearson} S.~G.,  {McCaughrean} M.~J.,  2023, \mn@doi [arXiv e-prints] {10.48550/arXiv.2310.01231}, \href {https://ui.adsabs.harvard.edu/abs/2023arXiv231001231P} {p. arXiv:2310.01231}

\bibitem[\protect\citeauthoryear{{Picogna}, {Ercolano}  \& {Espaillat}}{{Picogna} et~al.}{2021}]{Picogna21}
{Picogna} G.,  {Ercolano} B.,   {Espaillat} C.~C.,  2021, \mn@doi [\mnras] {10.1093/mnras/stab2883}, \href {https://ui.adsabs.harvard.edu/abs/2021MNRAS.508.3611P} {508, 3611}

\bibitem[\protect\citeauthoryear{{Pierens} \& {Nelson}}{{Pierens} \& {Nelson}}{2013}]{Pierens13}
{Pierens} A.,  {Nelson} R.~P.,  2013, \mn@doi [\aap] {10.1051/0004-6361/201321777}, \href {https://ui.adsabs.harvard.edu/abs/2013A&A...556A.134P} {556, A134}

\bibitem[\protect\citeauthoryear{{Pinte}, {Dent}, {M{\'e}nard}, {Hales}, {Hill}, {Cortes}  \& {de Gregorio-Monsalvo}}{{Pinte} et~al.}{2016}]{Pinte16}
{Pinte} C.,  {Dent} W.~R.~F.,  {M{\'e}nard} F.,  {Hales} A.,  {Hill} T.,  {Cortes} P.,   {de Gregorio-Monsalvo} I.,  2016, \mn@doi [\apj] {10.3847/0004-637X/816/1/25}, \href {https://ui.adsabs.harvard.edu/abs/2016ApJ...816...25P} {816, 25}

\bibitem[\protect\citeauthoryear{{Poon}, {Nelson}  \& {Coleman}}{{Poon} et~al.}{2021}]{Poon21}
{Poon} S. T.~S.,  {Nelson} R.~P.,   {Coleman} G. A.~L.,  2021, \mn@doi [\mnras] {10.1093/mnras/stab1466}, \href {https://ui.adsabs.harvard.edu/abs/2021MNRAS.505.2500P} {505, 2500}

\bibitem[\protect\citeauthoryear{{Portegies Zwart} \& {Hochart}}{{Portegies Zwart} \& {Hochart}}{2023}]{PortegiesZwart24}
{Portegies Zwart} S.,  {Hochart} E.,  2023, \mn@doi [arXiv e-prints] {10.48550/arXiv.2312.04645}, \href {https://ui.adsabs.harvard.edu/abs/2023arXiv231204645P} {p. arXiv:2312.04645}

\bibitem[\protect\citeauthoryear{{Pr{\v{s}}a} et~al.,}{{Pr{\v{s}}a} et~al.}{2011}]{Prsa11}
{Pr{\v{s}}a} A.,  et~al., 2011, \mn@doi [\aj] {10.1088/0004-6256/141/3/83}, \href {https://ui.adsabs.harvard.edu/abs/2011AJ....141...83P} {141, 83}

\bibitem[\protect\citeauthoryear{{Qiao}, {Coleman}  \& {Haworth}}{{Qiao} et~al.}{2023}]{Qiao23}
{Qiao} L.,  {Coleman} G. A.~L.,   {Haworth} T.~J.,  2023, \mn@doi [\mnras] {10.1093/mnras/stad944}, \href {https://ui.adsabs.harvard.edu/abs/2023MNRAS.522.1939Q} {522, 1939}

\bibitem[\protect\citeauthoryear{{Raghavan} et~al.,}{{Raghavan} et~al.}{2010}]{Raghavan10}
{Raghavan} D.,  et~al., 2010, \mn@doi [\apjs] {10.1088/0067-0049/190/1/1}, \href {https://ui.adsabs.harvard.edu/abs/2010ApJS..190....1R} {190, 1}

\bibitem[\protect\citeauthoryear{{Rasio} \& {Ford}}{{Rasio} \& {Ford}}{1996}]{RasioFord1996}
{Rasio} F.~A.,  {Ford} E.~B.,  1996, \mn@doi [Science] {10.1126/science.274.5289.954}, \href {http://adsabs.harvard.edu/abs/1996Sci...274..954R} {274, 954}

\bibitem[\protect\citeauthoryear{{Reipurth} \& {Clarke}}{{Reipurth} \& {Clarke}}{2001}]{Reipurth01}
{Reipurth} B.,  {Clarke} C.,  2001, \mn@doi [\aj] {10.1086/321121}, \href {https://ui.adsabs.harvard.edu/abs/2001AJ....122..432R} {122, 432}

\bibitem[\protect\citeauthoryear{{Rosotti}}{{Rosotti}}{2023}]{Rosotti23}
{Rosotti} G.~P.,  2023, \mn@doi [\nar] {10.1016/j.newar.2023.101674}, \href {https://ui.adsabs.harvard.edu/abs/2023NewAR..9601674R} {96, 101674}

\bibitem[\protect\citeauthoryear{{Sch{\"a}fer}, {Yang}  \& {Johansen}}{{Sch{\"a}fer} et~al.}{2017}]{Schafer17}
{Sch{\"a}fer} U.,  {Yang} C.-C.,   {Johansen} A.,  2017, \mn@doi [\aap] {10.1051/0004-6361/201629561}, \href {https://ui.adsabs.harvard.edu/abs/2017A&A...597A..69S} {597, A69}

\bibitem[\protect\citeauthoryear{{Shakura} \& {Sunyaev}}{{Shakura} \& {Sunyaev}}{1973}]{Shak}
{Shakura} N.~I.,  {Sunyaev} R.~A.,  1973, \aap, \href {http://adsabs.harvard.edu/abs/1973A%26A....24..337S} {24, 337}

\bibitem[\protect\citeauthoryear{{Smullen}, {Kratter}  \& {Shannon}}{{Smullen} et~al.}{2016}]{Smullen16}
{Smullen} R.~A.,  {Kratter} K.~M.,   {Shannon} A.,  2016, \mn@doi [\mnras] {10.1093/mnras/stw1347}, \href {https://ui.adsabs.harvard.edu/abs/2016MNRAS.461.1288S} {461, 1288}

\bibitem[\protect\citeauthoryear{{Spergel} et~al.,}{{Spergel} et~al.}{2015}]{Spergel15}
{Spergel} D.,  et~al., 2015, arXiv e-prints, \href {https://ui.adsabs.harvard.edu/abs/2015arXiv150303757S} {p. arXiv:1503.03757}

\bibitem[\protect\citeauthoryear{{Standing} et~al.,}{{Standing} et~al.}{2023}]{Standing23}
{Standing} M.~R.,  et~al., 2023, \mn@doi [Nature Astronomy] {10.1038/s41550-023-01948-4}, \href {https://ui.adsabs.harvard.edu/abs/2023NatAs...7..702S} {7, 702}

\bibitem[\protect\citeauthoryear{{Sumi} et~al.,}{{Sumi} et~al.}{2011}]{Sumi11}
{Sumi} T.,  et~al., 2011, \mn@doi [\nat] {10.1038/nature10092}, \href {https://ui.adsabs.harvard.edu/abs/2011Natur.473..349S} {473, 349}

\bibitem[\protect\citeauthoryear{{Sumi} et~al.,}{{Sumi} et~al.}{2023}]{Sumi23}
{Sumi} T.,  et~al., 2023, \mn@doi [\aj] {10.3847/1538-3881/ace688}, \href {https://ui.adsabs.harvard.edu/abs/2023AJ....166..108S} {166, 108}

\bibitem[\protect\citeauthoryear{{Sutherland} \& {Fabrycky}}{{Sutherland} \& {Fabrycky}}{2016}]{Sutherland16}
{Sutherland} A.~P.,  {Fabrycky} D.~C.,  2016, \mn@doi [\apj] {10.3847/0004-637X/818/1/6}, \href {https://ui.adsabs.harvard.edu/abs/2016ApJ...818....6S} {818, 6}

\bibitem[\protect\citeauthoryear{{Theissen}, {Konopacky}, {Lu}, {Kim}, {Zhang}, {Hsu}, {Chu}  \& {Wei}}{{Theissen} et~al.}{2022}]{Theissen23}
{Theissen} C.~A.,  {Konopacky} Q.~M.,  {Lu} J.~R.,  {Kim} D.,  {Zhang} S.~Y.,  {Hsu} C.-C.,  {Chu} L.,   {Wei} L.,  2022, \mn@doi [\apj] {10.3847/1538-4357/ac3252}, \href {https://ui.adsabs.harvard.edu/abs/2022ApJ...926..141T} {926, 141}

\bibitem[\protect\citeauthoryear{{Thiabaud}, {Marboeuf}, {Alibert}, {Cabral}, {Leya}  \& {Mezger}}{{Thiabaud} et~al.}{2014}]{Thiabaud14}
{Thiabaud} A.,  {Marboeuf} U.,  {Alibert} Y.,  {Cabral} N.,  {Leya} I.,   {Mezger} K.,  2014, \mn@doi [\aap] {10.1051/0004-6361/201322208}, \href {http://adsabs.harvard.edu/abs/2014A%26A...562A..27T} {562, A27}

\bibitem[\protect\citeauthoryear{{Thiabaud}, {Marboeuf}, {Alibert}, {Leya}  \& {Mezger}}{{Thiabaud} et~al.}{2015}]{Thiabaud15}
{Thiabaud} A.,  {Marboeuf} U.,  {Alibert} Y.,  {Leya} I.,   {Mezger} K.,  2015, \mn@doi [\aap] {10.1051/0004-6361/201424868}, \href {https://ui.adsabs.harvard.edu/abs/2015A&A...574A.138T} {574, A138}

\bibitem[\protect\citeauthoryear{{Thun}, {Kley}  \& {Picogna}}{{Thun} et~al.}{2017}]{Thun17}
{Thun} D.,  {Kley} W.,   {Picogna} G.,  2017, \mn@doi [\aap] {10.1051/0004-6361/201730666}, \href {https://ui.adsabs.harvard.edu/abs/2017A&A...604A.102T} {604, A102}

\bibitem[\protect\citeauthoryear{{Trapman}, {Rosotti}, {Bosman}, {Hogerheijde}  \& {van Dishoeck}}{{Trapman} et~al.}{2020}]{Trapman20}
{Trapman} L.,  {Rosotti} G.,  {Bosman} A.~D.,  {Hogerheijde} M.~R.,   {van Dishoeck} E.~F.,  2020, \mn@doi [\aap] {10.1051/0004-6361/202037673}, \href {https://ui.adsabs.harvard.edu/abs/2020A&A...640A...5T} {640, A5}

\bibitem[\protect\citeauthoryear{{Veras} \& {Raymond}}{{Veras} \& {Raymond}}{2012}]{Veras12}
{Veras} D.,  {Raymond} S.~N.,  2012, \mn@doi [\mnras] {10.1111/j.1745-3933.2012.01218.x}, \href {https://ui.adsabs.harvard.edu/abs/2012MNRAS.421L.117V} {421, L117}

\bibitem[\protect\citeauthoryear{{Villenave} et~al.,}{{Villenave} et~al.}{2020}]{Villenave20}
{Villenave} M.,  et~al., 2020, \mn@doi [\aap] {10.1051/0004-6361/202038087}, \href {https://ui.adsabs.harvard.edu/abs/2020A&A...642A.164V} {642, A164}

\bibitem[\protect\citeauthoryear{{Villenave} et~al.,}{{Villenave} et~al.}{2022}]{Villenave22}
{Villenave} M.,  et~al., 2022, \mn@doi [\apj] {10.3847/1538-4357/ac5fae}, \href {https://ui.adsabs.harvard.edu/abs/2022ApJ...930...11V} {930, 11}

\bibitem[\protect\citeauthoryear{{Wang}, {Kouwenhoven}, {Zheng}, {Church}  \& {Davies}}{{Wang} et~al.}{2015}]{Wang15}
{Wang} L.,  {Kouwenhoven} M.~B.~N.,  {Zheng} X.,  {Church} R.~P.,   {Davies} M.~B.,  2015, \mn@doi [\mnras] {10.1093/mnras/stv542}, \href {https://ui.adsabs.harvard.edu/abs/2015MNRAS.449.3543W} {449, 3543}

\bibitem[\protect\citeauthoryear{{Wang}, {Perna}  \& {Zhu}}{{Wang} et~al.}{2023}]{Wang24}
{Wang} Y.,  {Perna} R.,   {Zhu} Z.,  2023, \mn@doi [arXiv e-prints] {10.48550/arXiv.2310.06016}, \href {https://ui.adsabs.harvard.edu/abs/2023arXiv231006016W} {p. arXiv:2310.06016}

\bibitem[\protect\citeauthoryear{{Weidenschilling} \& {Marzari}}{{Weidenschilling} \& {Marzari}}{1996}]{Weidenschilling96}
{Weidenschilling} S.~J.,  {Marzari} F.,  1996, \mn@doi [\nat] {10.1038/384619a0}, \href {https://ui.adsabs.harvard.edu/abs/1996Natur.384..619W} {384, 619}

\bibitem[\protect\citeauthoryear{{Whitworth} \& {Zinnecker}}{{Whitworth} \& {Zinnecker}}{2004}]{Whitworth04}
{Whitworth} A.~P.,  {Zinnecker} H.,  2004, \mn@doi [\aap] {10.1051/0004-6361:20041131}, \href {https://ui.adsabs.harvard.edu/abs/2004A&A...427..299W} {427, 299}

\bibitem[\protect\citeauthoryear{{Winn} \& {Fabrycky}}{{Winn} \& {Fabrycky}}{2015}]{Winn15}
{Winn} J.~N.,  {Fabrycky} D.~C.,  2015, \mn@doi [\araa] {10.1146/annurev-astro-082214-122246}, \href {https://ui.adsabs.harvard.edu/abs/2015ARA&A..53..409W} {53, 409}

\bibitem[\protect\citeauthoryear{{Winter}, {Clarke}, {Rosotti}, {Ih}, {Facchini}  \& {Haworth}}{{Winter} et~al.}{2018}]{Winter18}
{Winter} A.~J.,  {Clarke} C.~J.,  {Rosotti} G.,  {Ih} J.,  {Facchini} S.,   {Haworth} T.~J.,  2018, \mn@doi [\mnras] {10.1093/mnras/sty984}, \href {https://ui.adsabs.harvard.edu/abs/2018MNRAS.478.2700W} {478, 2700}

\bibitem[\protect\citeauthoryear{{Winter}, {Haworth}, {Coleman}  \& {Nayakshin}}{{Winter} et~al.}{2022}]{Winter22}
{Winter} A.~J.,  {Haworth} T.~J.,  {Coleman} G. A.~L.,   {Nayakshin} S.,  2022, \mn@doi [\mnras] {10.1093/mnras/stac1564}, \href {https://ui.adsabs.harvard.edu/abs/2022MNRAS.515.4287W} {515, 4287}

\bibitem[\protect\citeauthoryear{{van Altena}, {Lee}, {Lee}, {Lu}  \& {Upgren}}{{van Altena} et~al.}{1988}]{VanAltena88}
{van Altena} W.~F.,  {Lee} J.~T.,  {Lee} J.~F.,  {Lu} P.~K.,   {Upgren} A.~R.,  1988, \mn@doi [\aj] {10.1086/114772}, \href {https://ui.adsabs.harvard.edu/abs/1988AJ.....95.1744V} {95, 1744}

\makeatother
\end{thebibliography}

\label{lastpage}
\end{document}